\documentclass[useAMS,usenatbib]{mnras}

\usepackage{graphicx}
\usepackage[T1]{fontenc}
\usepackage{aecompl}
\usepackage[]{amsmath}
\usepackage{color}
\usepackage{xspace}

\newcommand{\hbtp}{\textsc{hbt$+$}\xspace}

\newcommand{\msun}{\mathrm{M}_\odot}
\graphicspath{{figures/}}

\newcommand{\mytab}{\begin{table}[htb]}
\newcommand{\myfig}{\begin{figure}[htbp]}

\newcommand{\ud}{\mathrm{d}}

\def\lsim{ \lower .75ex \hbox{$\sim$} \llap{\raise .27ex \hbox{$<$}} }

\title[]{What to expect from dynamical modelling of galactic haloes II:
the spherical Jeans equation
}

\author[Wang et al.]{Wenting Wang$^{1}$\thanks{wenting.wang@ipmu.jp}, Jiaxin Han$^{1}$\thanks{jiaxin.han@ipmu.jp}, Shaun Cole$^{2}$,
Surhud More$^{1}$, Carlos Frenk$^{2}$,\newauthor Matthieu Schaller$^{2}$\\
  {}$^{1}$Kavli IPMU (WPI), UTIAS, The University of Tokyo, Kashiwa, Chiba 277-8583, Japan \\
  $^2$ Institute for Computational Cosmology, University of Durham, South Road, Durham, DH1 3LE, UK
}

\begin{document}

\maketitle

\begin{abstract}
The spherical Jeans equation (SJE) is widely used in dynamical modelling of the Milky 
Way (MW) halo potential. We use haloes and galaxies from the cosmological Millennium-II 
simulation and hydrodynamical APOSTLE simulations to investigate the performance 
of the SJE in recovering the underlying mass profiles of MW mass haloes. The best-fitting 
halo mass and concentration parameters scatter by 25\% and 40\% around their input values, 
respectively, when dark matter particles are used as tracers. This scatter becomes as 
large as a factor of 3 when using star particles instead. This is significantly larger 
than the estimated statistical uncertainty associated with the use of the SJE. The existence 
of correlated phase-space structures that violate the steady state assumption of the SJE as 
well as non-spherical geometries are the principal sources of the scatter. Binary haloes 
show larger scatter because they are more aspherical in shape and have a more perturbed 
dynamical state. Our results confirm the previous study of Wang et al. (2017) that the 
number of independent phase-space structures sets an intrinsic \emph{limiting precision} 
on dynamical inferences based on the steady state assumption. Modelling with a 
radius-independent velocity anisotropy, or using tracers within a limited outer radius, 
result in significantly larger scatter, but the ensemble-averaged measurement over the 
whole halo sample is approximately unbiased.
\end{abstract}   


\section{Introduction}
\label{sec:intro}

Our galaxy, the Milky Way (MW), provides a wealth of valuable information on the nature of 
the dark matter and the physics of galaxy formation. Many important inferences, however, depend 
on the precision with which the mass of its dark matter halo can be estimated. For example, 
the ``too big to fail'' problem claims that the structure of the most massive dark matter 
subhaloes predicted by $\Lambda$CDM simulations of MW-like hosts are inconsistent  with the 
structure of the classical dwarf satellites observed around the MW \citep{2011MNRAS.415L..40B, 
2012MNRAS.425.2817F}. The number of massive subhaloes in these simulations, which are 
inconsistent with the observed structure, is sensitive to the assumed MW halo mass, and 
the problem would disappear if the MW halo mass is sufficiently small 
\citep[$<1\times10^{12}\msun$;][]{2012MNRAS.424.2715W,2014MNRAS.445.1820C}.

There are many different approaches to measuring the underlying potential of the MW. A 
brief summary of previous results can be found in \cite{2015MNRAS.453..377W}. A few more 
recent measurements include \citet{2016MNRAS.463.2623H}, \cite{2017arXiv170800170A}, 
\cite{2017MNRAS.465...76M}, \cite{2017MNRAS.468.3428P} and \cite{2017MNRAS.467.1844R}. 
The inference of the halo mass from such observations unavoidably involves various model 
assumptions, which are often not entirely justified or tested on realistic numerical 
simulations.

In a series of previous studies, we have examined the validity of a few such model assumptions, 
by applying the relevant method to simulated dark matter haloes and galaxies for which the 
underlying potentials are known. In \cite{2015MNRAS.453..377W}, we tested the method of 
fitting a given model distribution function to the observed radial and velocity distribution 
of dynamical tracers such as halo stars, globular clusters and luminous satellite galaxies 
\citep[e.g.,][]{1999MNRAS.310..645W, 2003A&A...397..899S, 2008MNRAS.388..815W, 2009MNRAS.399..812W, 
2010MNRAS.408.2442W, 2012MNRAS.424L..44D, 2015ApJ...806...54E, 2017ApJ...850..116L}. Strong 
deviations between measured and true halo parameters were found. Multiple factors are 
responsible for the discrepancy, including deviations from the adopted functional form 
of the model potential, deviations from spherical symmetry, violations of the form of 
the distribution function and violations of the steady state assumption for the dynamical 
tracers.

\cite{han2016a} developed the orbital probability 
distribution method (oPDF) which involves only two model assumptions: (1) 
the potential is spherical  (2) the system is in a steady state. The oPDF 
method expresses the steady-state solution of the collisionless Boltzmann 
equation as a microscopic equilibrium distribution function, from which 
one can predict the radial distribution of tracers given a model potential 
and the observed positions and velocities of the tracers. The predicted radial 
distribution is then compared with the observed distribution to derive the 
best-fitting potential. Although the method only works with six dimensional phase 
space data in its current form, it involves only the most basic model 
assumptions, which enables us to understand the uncertainties from these 
assumptions in a focused way. \cite{han2016b} and \cite{2017MNRAS.470.2351W} 
applied this method to large samples of dark matter haloes and galaxies in 
the Aquarius simulations \citep{2010MNRAS.406..744C}, the Millennium-II simulation 
\citep{2009MNRAS.398.1150B} and the hydrodynamical APOSTLE simulations 
\citep[A Project of Simulations of The Local Environment;][]{2016MNRAS.457..844F,
2016MNRAS.457.1931S}. In their analysis, the true potential profiles in 
the simulation are extracted as model templates, and thus the results are 
free from uncertainties due to imperfections in the assumed potential 
profile.

It has been found that violations of the two model assumptions above can lead to 
about 25\% uncertainty in the halo mass when dark matter particles are used as 
tracers. This uncertainty increases to 200-300\% when stars are used as 
tracers~\citep{2017MNRAS.470.2351W}. This uncertainty cannot be trivially 
decreased by increasing the tracer sample size, reflecting a \emph{limiting 
precision} linked to the intrinsic number of phase-independent particles in 
each halo. This intrinsic number is smaller than the actual size of the tracer 
sample, due to correlations in the phase space coordinates of the tracer particles 
that violate the steady state assumption. In particular, \cite{2017MNRAS.470.2351W} 
explicitly demonstrate that an effective sample size estimated from the distribution 
of streams correlates with the amplitude of the uncertainty in the best fits 
inferred using  the oPDF method. 

In this paper, we further test the approach that uses the spherical Jeans equation 
(hereafter SJE) to infer the underlying mass profile or circular velocity curve. 
SJE has been widely used to measure the halo circular velocity of MW, $V_\mathrm{circ}$, 
from the radial velocity dispersion of tracers, $\sigma_r(r)$ \citep[e.g.][]{2005MNRAS.364..433B,
2008ApJ...684.1143X,2010MNRAS.406..264W,2010ApJ...720L.108G,2012ApJ...761...98K,2014ApJ...794...59K,
2017arXiv170800170A}. Both the oPDF and the Jeans equation are derived from the collisionless 
Boltzmann equation. The SJE widely used in the literature also depends on the 
assumptions of steady state tracers and a spherical potential. 

Applying the SJE requires the tracer velocity anisotropy, $\beta$, and density 
profiles, $\rho_\ast$, to be known. In reality, $\beta$ and $\rho_\ast$ are often 
not available and have to be assumed or marginalised over. We first use the full set 
of simulation data to calculate tracer properties, so that any uncertainty from 
the unknown velocity anisotropy and density profiles is not a concern. This allows 
us mainly to check uncertainties from the steady state and the spherical assumption, 
which then enables direct comparisons with \cite{han2016b} and \cite{2017MNRAS.470.2351W}. 
In the end, we also investigate what happens if $\beta$ is modelled as a constant, 
either with an assumed value or as a free parameter. We also discuss the result if 
only tracers within a given radial range are used.

After we have finalised this work, \citet{2018MNRAS.475.4434K} published a related 
study that tested the SJE in recovering the mass profile from 10 to 100~kpc. While 
we have thoroughly studied how well the potential profile can be recovered from different 
dynamical models in a series of previous studies~\citep{2015MNRAS.453..377W,han2016a,
han2016b,2017MNRAS.470.2351W}, in this work we focus on presenting results on the 
recovered virial mass and concentration parameters which are of more cosmological 
interests. We only briefly revisit the recovered potential profile in the context of 
SJE that shows consistent behaviour with our previous findings. Compared with 
\citet{2018MNRAS.475.4434K}, our halo sample is much larger and our analysis of the 
source of uncertainties is more complete and thorough.

\section{Simulations and Tracers}
\label{sec:mock}

We use the same data sets as previously analysed by \cite{han2016a} and \cite{2017MNRAS.470.2351W}.  
Our analysis involves 120 ideal haloes, more than 1000 isolated and binary haloes selected from 
cosmological N-body simulation and 24 galaxies (or 12 pairs) in hydro-dynamical simulations of 
the Local Group. Further information can be found in the remainder of this section. Throughout 
this paper, we do not include particles belonging to subhaloes in our tracer sample. A thorough 
discussion of the further influence of subhaloes can be found in \cite{han2016b}.

\subsection{Ideal tracers}
\label{sec:idealhalo}
In order to test the SJE method in the ideal case, we first generate a steady-state 
system of tracers according to the probability distribution $\ud P(r,v)=f(E)L^{-2
\beta}\ud^3r\ud^3v$ used in \cite{2015MNRAS.453..377W}. The detailed form of $f(E)$ 
is specified by assuming a Navarro-Frenk-White\citep[NFW][]{Navarro_1996,Navarro_1997} 
potential and requiring the tracer density profile to be a double-power law.
The complete form of this distribution function and its derivation can be found in 
Equation 12 of \cite{2015MNRAS.453..377W} and the corresponding section. It describes 
a steady-state spherical system of tracers inside an NFW halo. The model has six parameters, 
including the mass, $M$, and concentration, $c$, of the NFW halo, the tracer velocity 
anisotropy, $\beta$, the double power law slopes of the tracer density profile 
$\alpha$ and $\gamma$, as well as the pivot radius of the tracers, $r_c$. Their values 
are chosen to best match the distribution of mock stars inside a MW sized halo in the 
Aquarius simulation \citep{2010MNRAS.406..744C, 2015MNRAS.446.2274L}, with 
$M=1.83\times 10^{12}\msun$, $c=16.2$, $\beta=0.715$, $r_c=69$~kpc, 
$\alpha=2.3$, $\gamma=7.47$. Tracer particles are generated between 10 and 1000 kpc 
in radius. We generate 120 samples, and each of them contains 4500 particles. We will 
call them ideal tracers.

\subsection{Millennium~II}
\label{sec:MRIIhalo}
A large sample of more realistic haloes are selected from the 
Millennium-II Simulation \citep[][hereafter MRII]{2009MNRAS.398.1150B}. 
MRII is a dark matter only simulation with a box size of 100~$h^{-1}$Mpc 
and a particle mass of $6.9\times10^6h^{-1}\mathrm{M_\odot}$. The 
cosmological parameters are those from the first year WMAP result 
\citep[][$\Omega_\mathrm{m}=0.25$, $\Omega_\Lambda=0.75$, $h=0.73$, 
$n=1$ and $\sigma_8=0.9$]{2003ApJS..148..175S}. 

To select haloes suitable for our analysis, we first identify a 
parent sample of haloes whose masses are analogous to MW, i.e., 
$0.5\times10^{12}<M_{200}<2.5\times10^{12}~\msun$ \footnote{We use $M_{200}$ 
to denote the mass of a spherical region with 
mean density equal to 200 times the critical density, $\rho_\mathrm{crit}$ 
of the Universe. The radius of the spherical region will be defined as 
the halo virial radius throughout the paper, denoted as $R_{200}$.}. 
Starting from these haloes we further select a sample of isolated haloes 
and a sample of binary ones. For isolated haloes, we require that all companions 
within a sphere of 2~Mpc should be at least one order of magnitude smaller 
in $M_{200}$. For binary haloes, we require the two haloes to be separated 
by a distance of 500 to 1000~kpc, to mimic the configuration of the MW and M31 
system. In addition, for a sphere centred on the mid-point of the two 
haloes and with a radius of 1.25~Mpc, all haloes within the sphere 
should be less massive than the smaller of the binary. In the end we 
have 658 isolated haloes and 336 binary haloes (or 168 pairs). Each 
halo contains about $10^5$ dark matter particles that are within $R_{200}$ 
and not bound to any substructure.

\subsection{The APOSTLE simulations}
\label{sec:LGhalo}
The APOSTLE~\citep[A Project of Simulations of The Local Environment][]{2016MNRAS.457..844F,2016MNRAS.457.1931S}
simulation is a set of zoomed hydrodynamical simulations of Local Group-like haloes in a 
$\Lambda$CDM universe, run using the same simulation code and parameters as the 
\textsc{eagle}~\citep{2015MNRAS.450.1937C, 2015MNRAS.446..521S} simulation. It consists of 12 
realisations, each representing a pair of galaxies analogous to the MW and M31 system. 
The underlying cosmology of APOSTLE is that of WMAP7 \citep[][$\Omega_\mathrm{m}=0.272$, 
$\Omega_\Lambda=0.728$, $h=0.704$, $n=0.967$ and $\sigma_8=0.81$]{Komatsu2011}. Each 
realisation is simulated at three different resolutions. The particle mass of the 
lowest resolution run is comparable to the intermediate resolution \textsc{eagle} run. 
The mass resolution of intermediate and high runs are higher than the lowest resolution 
runs by factors of 12 and 144 respectively, but the high resolution runs are 
not yet complete for all 12 volumes. For our analysis, we choose to use the 
suite of intermediate resolution runs. Each galaxy in the intermediate level 
contains about $\sim 10^4$ to $\sim 10^5$ star particles in the stellar halo 
that are not bound to any satellites.

\section{Methodology}
\label{sec:method}
Assuming the Galactic halo is spherical and in a steady state, we can 
derive the the spherical Jeans equation (SJE; Binney and Tremaine 1987):
\begin{equation}
 \frac{1}{\rho_\ast}\frac{\ud(\rho_\ast \sigma^2_{r,\ast})}{\ud r}+\frac{2\beta \sigma^2_{r,\ast}}{r}=-\frac{\ud \phi}{\ud r}=-\frac{V_c^2}{r}.
 \label{eqn:pot}
\end{equation}
We measure the radial velocity dispersion of tracers, $\sigma_{r,\ast}$, their velocity anisotropy, 
$\beta$, and the radial profile, $\rho_\ast$ from the simulations. Thus the potential 
gradient, or the rotation curve of the halo can be directly inferred from Equation~\ref{eqn:pot}. 

To obtain parameters of the halo, we first fit an NFW potential gradient,
\begin{equation}
 \frac{\ud \phi_\mathrm{NFW}}{\ud r}=-4\pi G \rho_s r_s [\frac{1}{(r/r_s)(1+r/r_s)}-\frac{1}{(r/r_s)^2}\log(1+\frac{r}{r_s})],
\end{equation}
to the Jeans inferred potential by varying two parameters, $\rho_s$ and $r_s$. The parameter $r_s$ 
is the radius where the effective logarithmic slope of the halo density profile is $-2$. These parameters
can be converted to the halo mass, $M_{200}$, and concentration parameter, $c_{200}=R_{200}/r_s$, 
through the following relations:
\begin{eqnarray}
 \rho_s=\frac{200\rho_\mathrm{crit}c_{200}}{3[\log(1+c_{200}-\frac{c_{200}}{1+c_{200}})]},
 \label{eqn:rel1}\\
 r_s=\left( \frac{3M_{200}}{800\pi \rho_\mathrm{crit} c_{200}^3} \right)^\frac{1}{3}.
 \label{eqn:rel2}
\end{eqnarray} 
We infer the best fit parameters by minimising
\begin{equation}
 \chi^2= {\bf d}^{\rm T} C_\mathrm{d}^{-1} {\bf d}
\end{equation}
where the vector ${\bf d}$ is given by
\begin{equation}
{\bf d} = \left({\bf \frac{\ud \phi_\mathrm{NFW}}{\ud r}-\frac{\ud \phi_\mathrm{jeans inferred}}{\ud r}}\right)\,.
\end{equation}
The data covariance matrix $C_\mathrm{d}$ is calculated from 100 bootstrap resamples generated with replacements 
preserving the sample size.

The minimisation of $\chi^2$ is achieved using the software \textsc{iminuit}, which is a python 
interface to the \textsc{minuit} function minimiser \citep{1975CoPhC..10..343J}. The statistical 
errors and covariance matrix of best-fitting parameters\footnote{The parameter covariance is not to 
be confused with the data covariance matrix $C_\mathrm{d}$.} are calculated from the Hessian matrix 
(i.e. gradient) of the $\chi^2$ with respect to the parameters.

The above approach enables us to focus on testing the steady state and spherical assumptions. 
However, it is a purely theoretical approach. In reality, the observable quantity is the tracer 
radial velocity dispersion, $\sigma_{r,\ast}$, and $\beta$ is often unknown. Assuming $\beta$ 
is constant, the solution to Equation~\ref{eqn:pot} reads
\begin{equation}
 \sigma^2_{r,\ast}(r)=\frac{1}{r^{2\beta}\rho_\ast(r)}\int_r^{\infty}\ud r' r'^{2\beta}\rho_\ast(r')\ud \phi/ \ud r ,
 \label{eqn:sigmar}
\end{equation}
subject to the boundary condition that lim$_{r \to \infty}r^{2\beta}\rho_\ast\sigma^2_{r,\ast}=0$ 
\citep[e.g.][]{2005MNRAS.364..433B,2012ApJ...761...98K}. To assess the practical application of the 
SJE, in Section~\ref{sec:discussion} we will also use Equation~\ref{eqn:sigmar} to fit the measured 
radial velocity dispersion profiles in the simulation, by treating $\beta$ as a radius-independent 
parameter.

\cite{han2016b} and \cite{2017MNRAS.470.2351W} have used both the NFW model profile and 
potential templates extracted from the true shape of potential profiles of haloes in the 
simulation. In this paper, we will focus on presenting results based on the NFW potential, 
because in practice it is not possible to know the true shape of the potential profile in 
advance and we have checked that the NFW model profile and the true potential templates 
give very similar levels of uncertainties in the best-fitting halo parameters 
\cite[see][]{2017MNRAS.470.2351W}. However, it has also been found by \cite{2017MNRAS.470.2351W} 
that deviations from the NFW model can cause a systematic bias, an underestimated $M_{200}$ 
and overestimated $c_{200}$ when tracers in the very inner halo are used. We also discuss this.

To obtain the true $c_{200}$ as a reference and compare with the best-fitting values, we adopt 
two approaches: (1) Directly fitting the NFW model to the true halo density profiles in the 
simulation and using the best-fitting $c_{200}$. (2) Finding the scale radius, $r_s$, where the 
logarithmic slope of the halo density profile equals -2 and estimating $c_{200}$ through 
$c_{200}=R_{200}/r_s$. $c_{200}$ calculated in these two ways shows less than 5\% difference, 
and hence uncertainties due to how the reference $c_{200}$ are defined are negligible.

\section{Ideal tracers and the statistical error of the fits}
\label{sec:ideal}

\begin{figure} 
\includegraphics[width=0.49\textwidth]{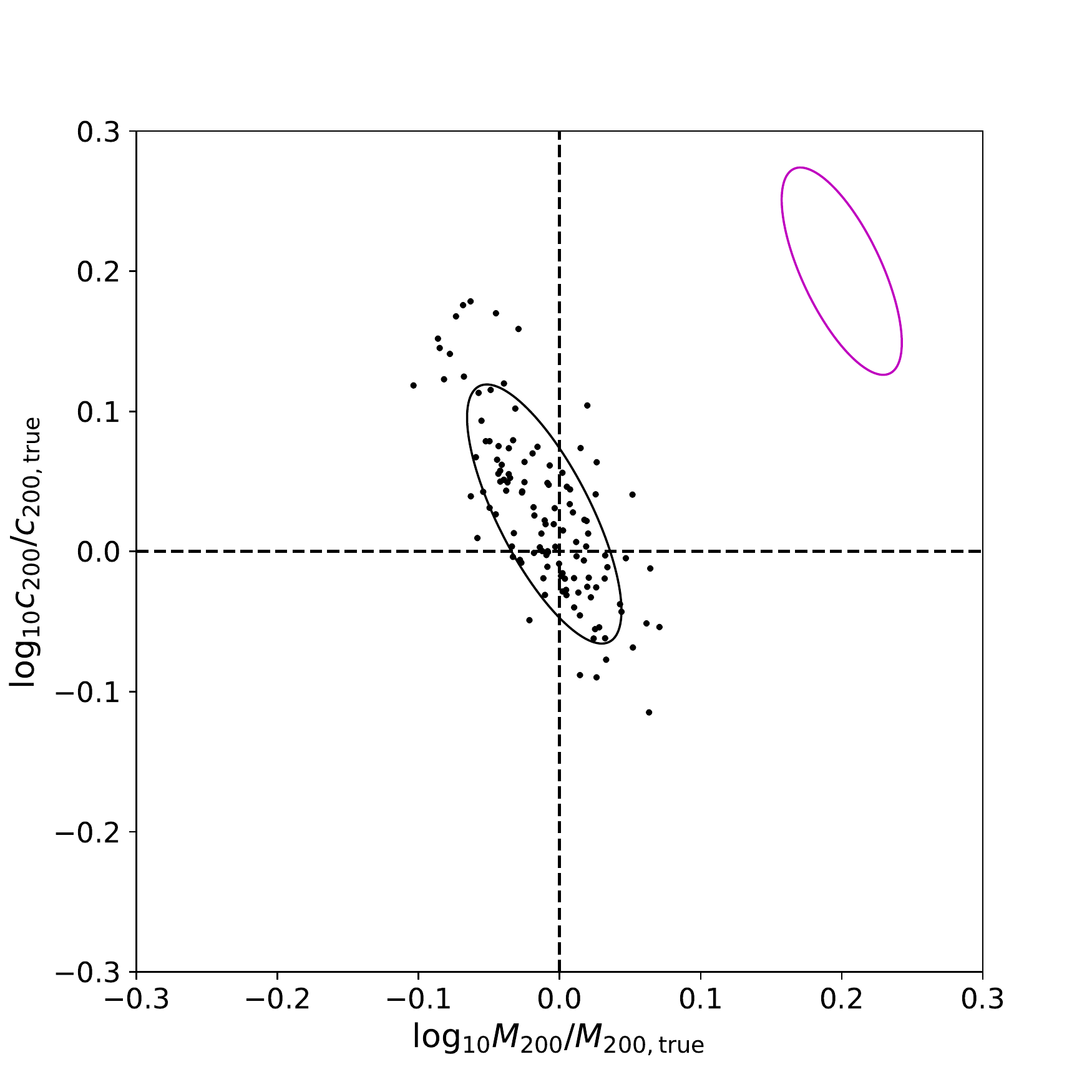}
\caption{Best-fitting halo mass ($x$-axis) and concentration ($y$-axis) in units of their true 
values, for 120 Monte-Carlo realisations of ideal haloes generated with a spherical NFW potential. 
Each dot represents the fit to one halo. Horizontal and vertical black dashed lines mark the equality 
between best-fitting and true parameters. The black ellipse marks the 1-$\sigma$ scatter of all 
the measurements. The magenta ellipse in the top right corner shows the average 1-$\sigma$ statistical 
error for a single halo.} 
\label{fig:2Dideal} 
\end{figure}

\begin{figure*} 
\includegraphics[width=0.49\textwidth]{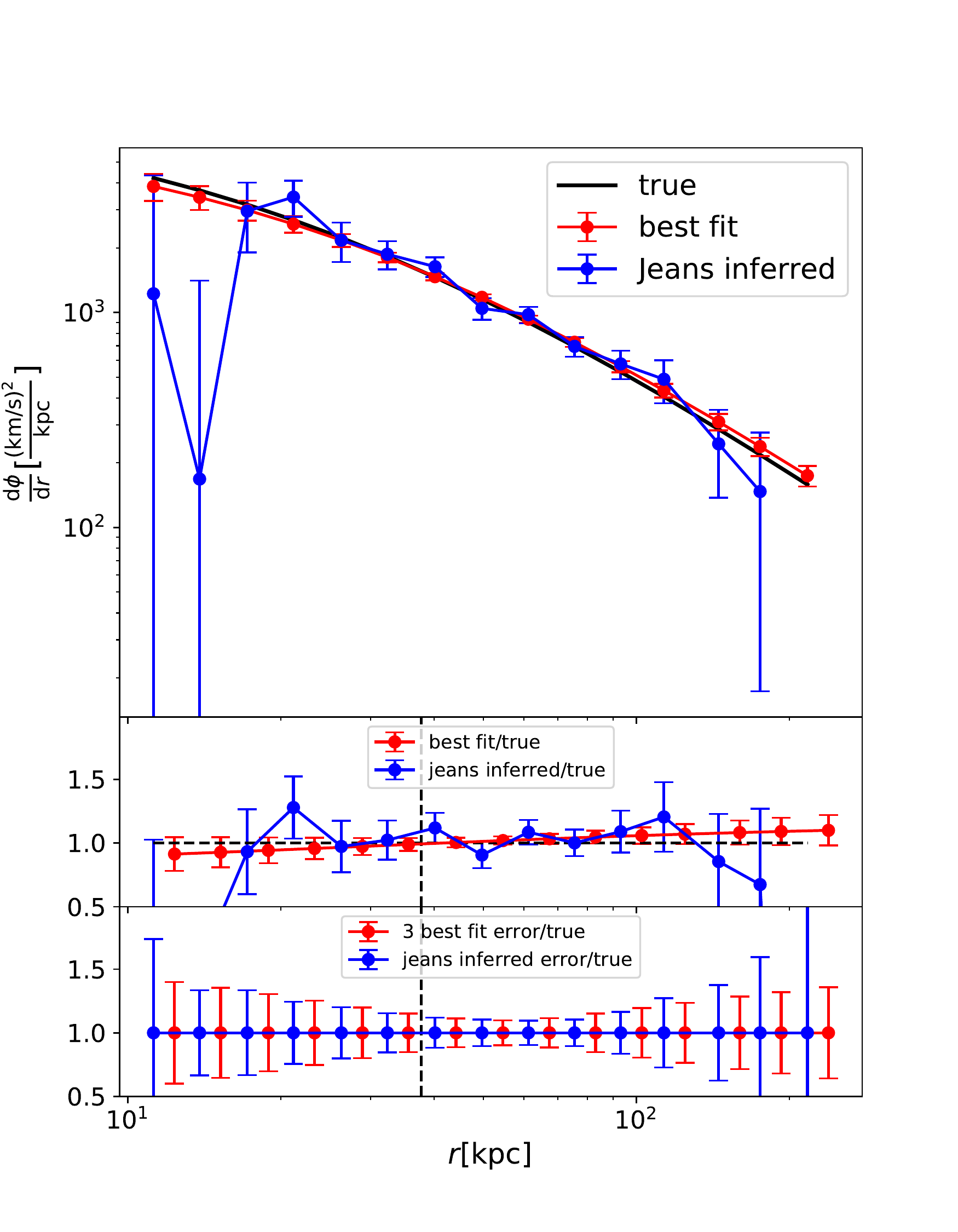}
\includegraphics[width=0.49\textwidth]{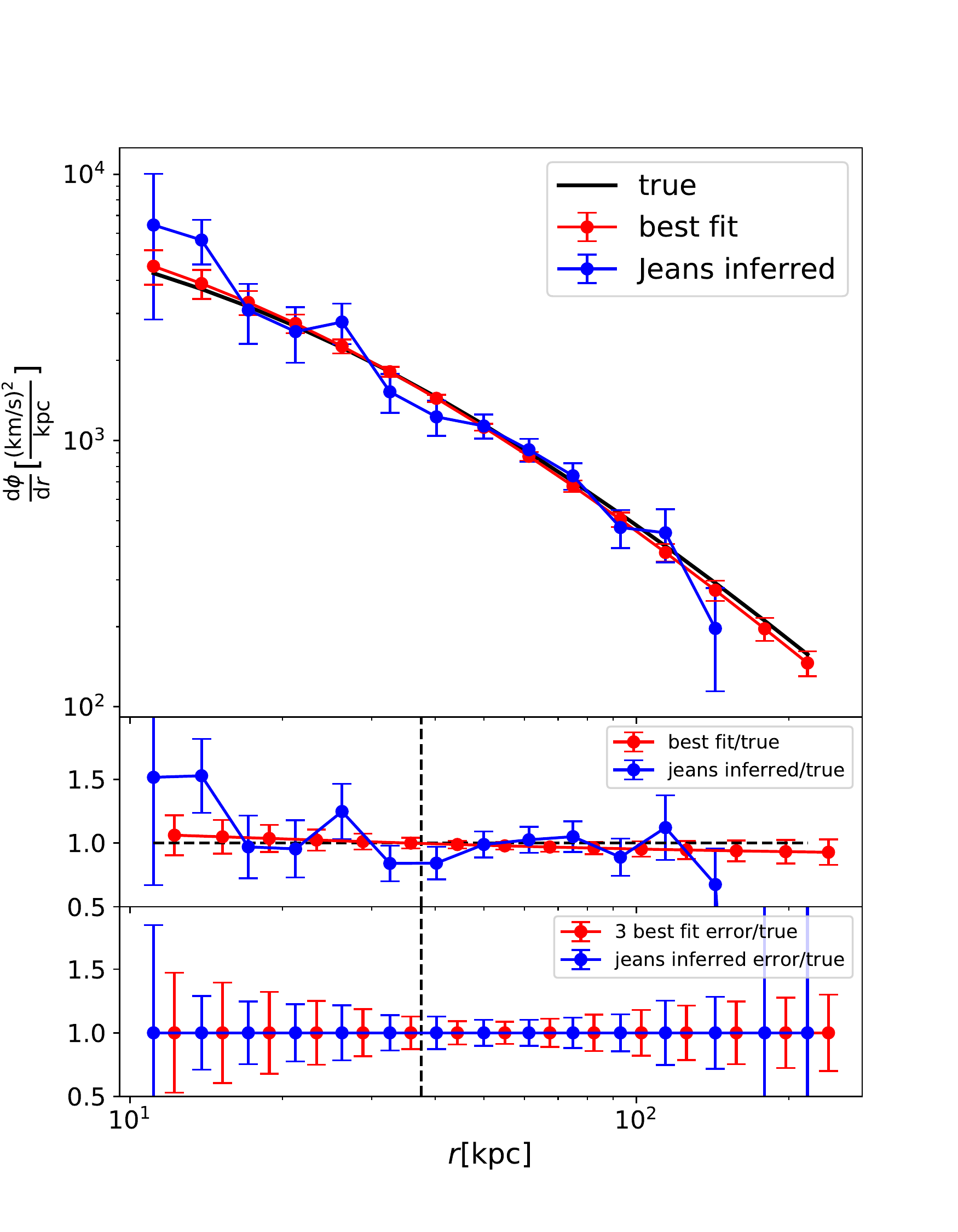}
\caption{Upper panels show comparisons of true halo potential gradient profiles in 
the simulation (black), Jeans profiles inferred through Equation~\ref{eqn:pot} (blue) 
and best-fitting profiles (red) for two randomly selected ideal haloes. Errors on the 
inferred profile are obtained through 100 bootstrap resamples of the full parent sample, 
while those on the best-fitting profile are computed from the covariance matrix of the 
best-fitting parameters. In the middle panel, the Jeans inferred profiles, the best-fitting 
profiles and the errors are all scaled by the true profile. The bottom panel is similar 
to the middle panel, except that each is scaled by its own profile to compare only the 
errors. In this panel, the errors on the best-fitting profiles are increased by a factor 
of 3, to make an easier comparison with those on the inferred profile. Vertical dashed 
lines mark the position of the median or half-mass radius for tracers. 
}
\label{fig:idealprof}
\end{figure*}

Ideal tracers in our analysis are steady state systems generated in a spherical and 
stationary NFW potential, and thus the SJE should be directly applicable. It is still 
interesting to apply our method to this system in order to test the performance and 
understand the error structure in the parameters. The best-fitting halo parameters 
using Equation~\ref{eqn:pot} are shown in Fig.~\ref{fig:2Dideal}. The statistical 
error size is {\it comparable} to the scatter in the best-fitting parameters 
(black ellipse in the middle). This is because our ideal tracer sample is free of 
any systematic uncertainty by construction and the scatter is dominated by statistical 
errors. We will show in the following sections that using more realistic tracers 
from cosmological simulations ends up with a much larger difference between 
statistical errors and the scatter in the best-fitting parameters.  

Similar to \cite{2017MNRAS.470.2351W} and \cite{han2016b}, the statistical error tends 
to align with a direction of anti-correlation between $M_{200}$ and $c_{200}$. In 
fact, the anti-correlation between halo mass and concentration (or other combinations 
of equivalent parameters) is commonly seen in dynamical modelling of the galactic 
potential \citep[e.g.][]{2014MNRAS.439.2678D,2014ApJ...794...59K,2015MNRAS.453..377W}, 
despite distinctions among different models. No matter how different the detailed 
approaches are, they all aim to fit the underlying potential profile, which can be 
well modelled by a double power law functional form. The parameter anti-correlation 
may partly arise from such a functional form, as we know power law fitting usually 
results in anti-correlations between the amplitude parameter and the shape parameter. 

The upper panels of Fig~\ref{fig:idealprof} show examples of true Jeans inferred and 
best-fitting potential gradient profiles for two ideal haloes. The differences can be 
seen more clearly in the middle panels, where we plot profiles normalised by the true 
one. The errors are scaled by the true profile and are thus relative errors. 
Over the whole radial range, deviations from the true profile are smaller than the 
errors, reflecting the fact that for ideal systems the uncertainties are properly 
modelled by statistical noise. 

To clearly show the radial variation of the errors, in the bottom panels we plot the error 
bars centred at the horizontal unity line. Both reveal a trend of being the smallest in the 
middle and largest on both smaller and larger scales. The smallest error occurs at a radius 
which is close but  not equal to the median or half-mass radius of tracers (vertical dashed 
line).

Given the smaller errors on such intermediate scales, the mass inside some intermediate 
radius can be constrained much better. We have checked that the 1-$\sigma$ uncertainty 
of the best-fitting mass within the median radius of tracers is smaller than that of 
$M_{200}$ by a factor of two in log-space. The better constraint of the total mass 
within the half light radius of tracers in dwarf MW satellite galaxies have been widely 
reported \citep[e.g.][]{2008ApJ...672..904P,2009ApJ...704.1274W,2010MNRAS.406.1220W}, though 
the adopted approaches and discussion are not the same. For example, \cite{2010MNRAS.406.1220W} 
provide the theoretical justification for why the mass within $r_3$, the radius where 
$\ud \log \rho_\ast/\ud \log r=-3$, is insensitive to the velocity anisotropy of tracers. 
For MW dwarf satellite galaxies, $r_3$ is close to the projected half-light radius. In our 
analysis, $\beta$ at all radii is known directly from the simulation,
and the better constrained mass within the median tracer radius is a reflection of 
the parameter anti-correlation. For parameter combinations along the anti-correlation 
direction, a larger estimated mass corresponding to a less concentrated profile, while a 
lower mass corresponding to a more concentrated halo. Thus one would naturally expect 
the mass profiles predicted by parameter combinations along the anti-correlation 
direction to cross with each other on some intermediate scale, i.e., the mass within 
an intermediate scale will have the least uncertainty. Interested readers can find more 
discussions in \cite{han2016a}. 

For Jeans inferred profiles, the smallest error does not have to occur around the 
median tracer radius, however. This is because for each radial bin, the error size 
is determined locally and does not depend on the overall radial range of the tracers. 
The fact that the errors in Fig.~\ref{fig:idealprof} are observed to be smallest 
near the median radius has to be connected to the particular radial distribution of 
the tracers. The tracers are most sparse on the largest scale, while the phase-space 
volume is smallest on small scales. This introduces larger variations on those two 
scales in general. Our current choice of model distribution function and radial 
ranges happen to lead to a smallest error near the median radius. 

\section{Tracers in realistic haloes}
Ideal tracers demonstrate that when the system is free of systematic sources of 
bias, our statistical error estimate correctly describes the scatter in best-fitting 
parameters. In this section we move on to use dark matter particles in MRII and 
star particles in APOSTLE as our more realistic tracers. We exclude particles 
in subhaloes\footnote{Particles in subhaloes are only excluded in the tracer 
samples. When calculating the true potential profile to extract true halo parameters, 
all particles are used.} and adopt an inner radius cut 
of 20~kpc for both simulations. The inner cut helps to avoid the central disc 
region in APOSTLE. 

For realistic tracers, there are three  sources of systematic errors: (1) Violation of the 
steady state assumption ($\sigma_\mathrm{steady}$) (2) Violation of the spherical assumption 
($\sigma_\mathrm{sph}$) (3) Violation of the assumed potential profile ($\sigma_\mathrm{NFW}$). 
Accordingly, the covariance matrix of the model parameters can be formally decomposed as 
a sum of the three plus statistical uncertainty ($\sigma_\mathrm{stat}$), 

\begin{equation}
{\bf C}_\mathrm{best-fit}=
{\bf C}_\mathrm{steady}+{\bf C}_\mathrm{sph}(+{\bf C}_\mathrm{NFW})+
{\bf C}_\mathrm{stat}.
\label{eqn:errors}
\end{equation}

\cite{2017MNRAS.470.2351W} showed that results based on true potential templates and 
the NFW profile give very similar uncertainty in best-fitting parameters if tracers within 
20~kpc are excluded. Therefore the effect of (3) is sub-dominant with respect to (1) and (2), 
and we have put ${\bf C}_{\rm NFW}$ in the bracket. These systematic sources of errors will 
be investigated in this section. If velocity anisotropy is unknown in the data, additional 
systematics may be introduced due to improper modelling of this component, and we postpone 
this discussion to section~\ref{sec:fixbeta}.

\subsection{Dark matter particles in MRII}
\label{sec:MRII}

\begin{figure*} 
\includegraphics[width=0.49\textwidth]{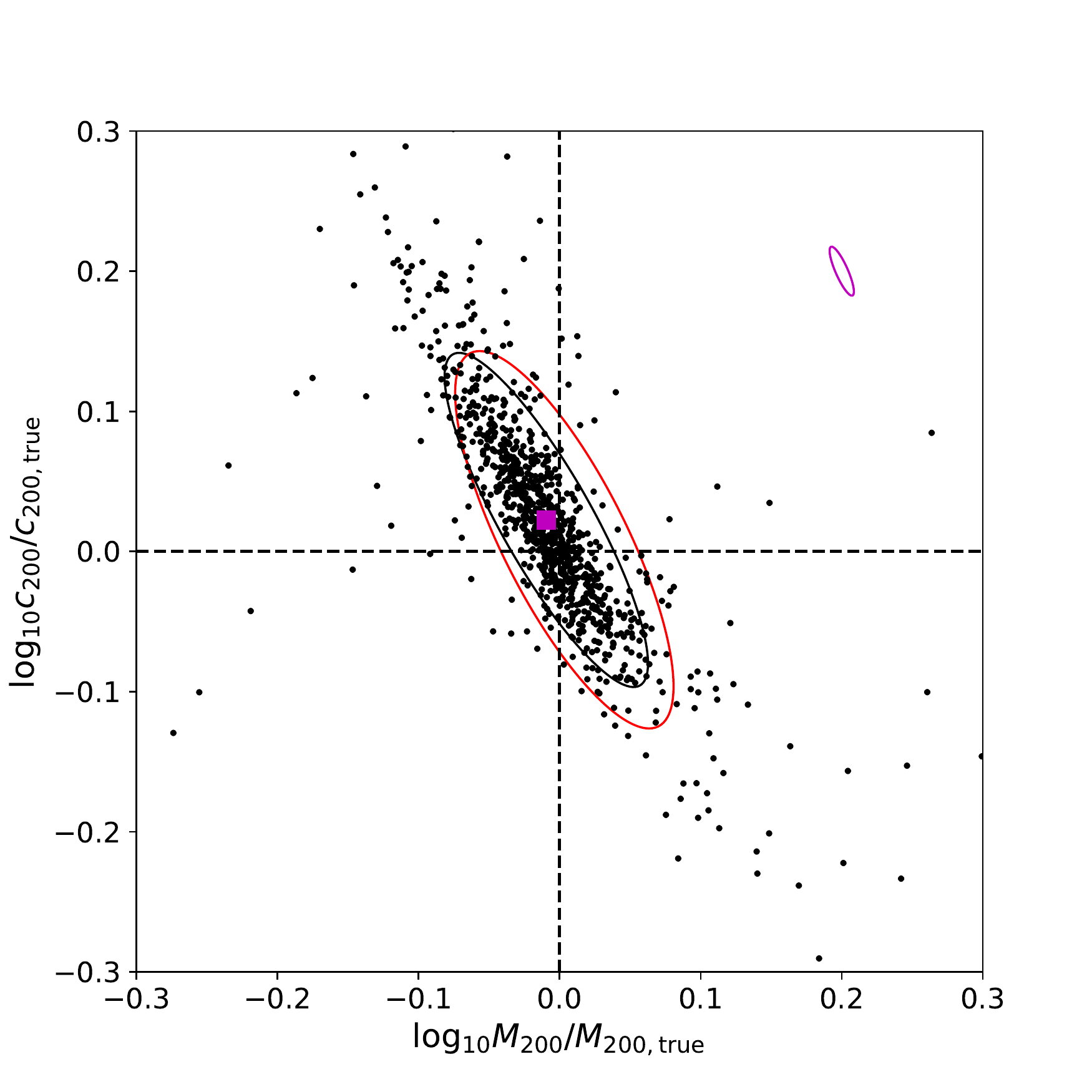}
\includegraphics[width=0.49\textwidth]{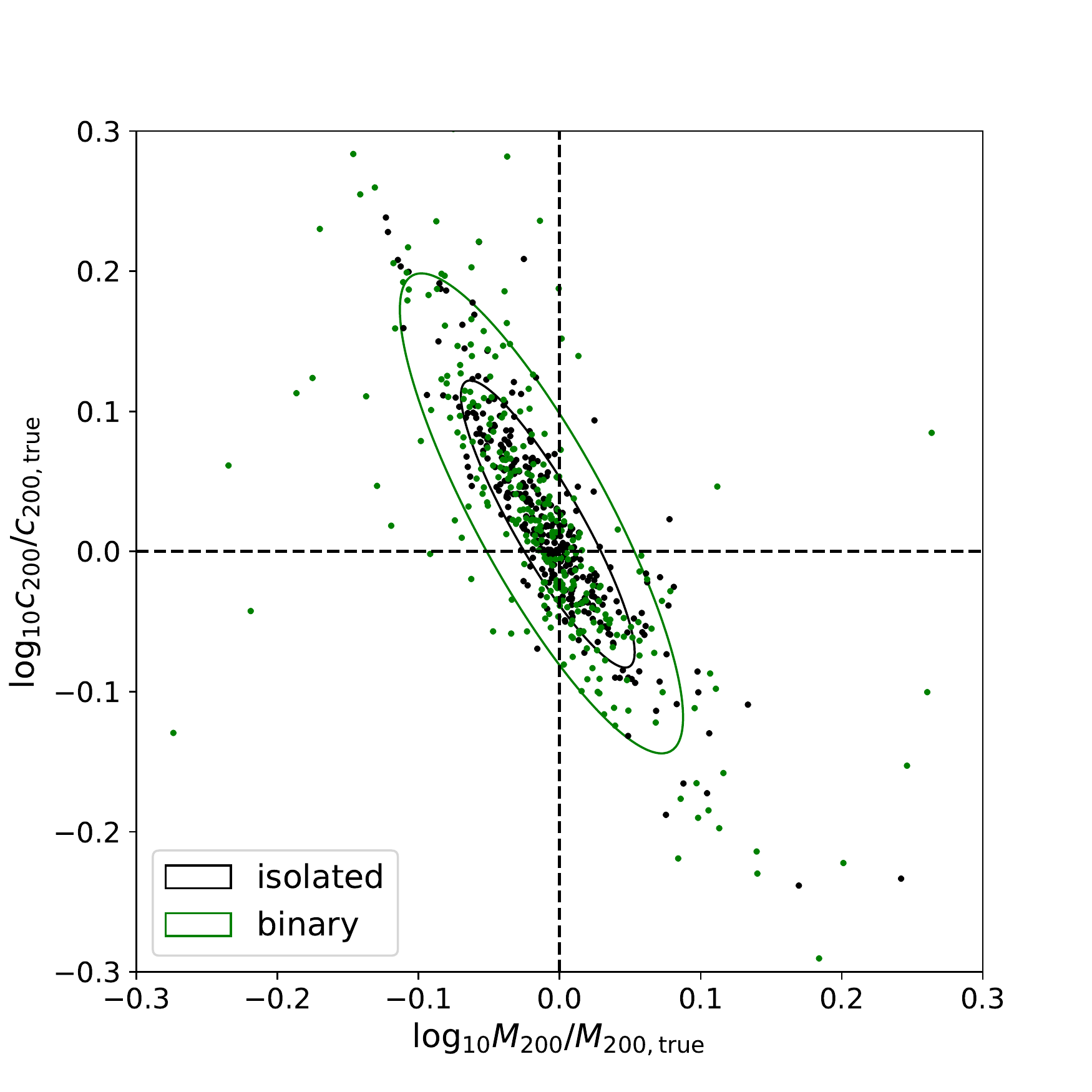}
\caption{{\bf Left:} Similar to Fig.~1, but based on all selected 
haloes in MRII. The black ellipse in the middle and the magenta ellipse 
in the top right show the average size of the uncertainty in the best-fitting parameters 
and the statistical errors. The red ellipse in the middle is a reproduction of 
the uncertainty in the best-fitting parameters in Fig.~1 of \citet{2017MNRAS.470.2351W} 
using the oPDF method. {\bf Right:} Best fits to binary (black) and isolated 
(green) haloes in MRII. Isolated haloes have been matched in mass to binary haloes 
to ensure the same halo mass distribution. } 
\label{fig:MRIIcomb} 
\end{figure*}

\begin{figure*} 
\includegraphics[width=0.49\textwidth]{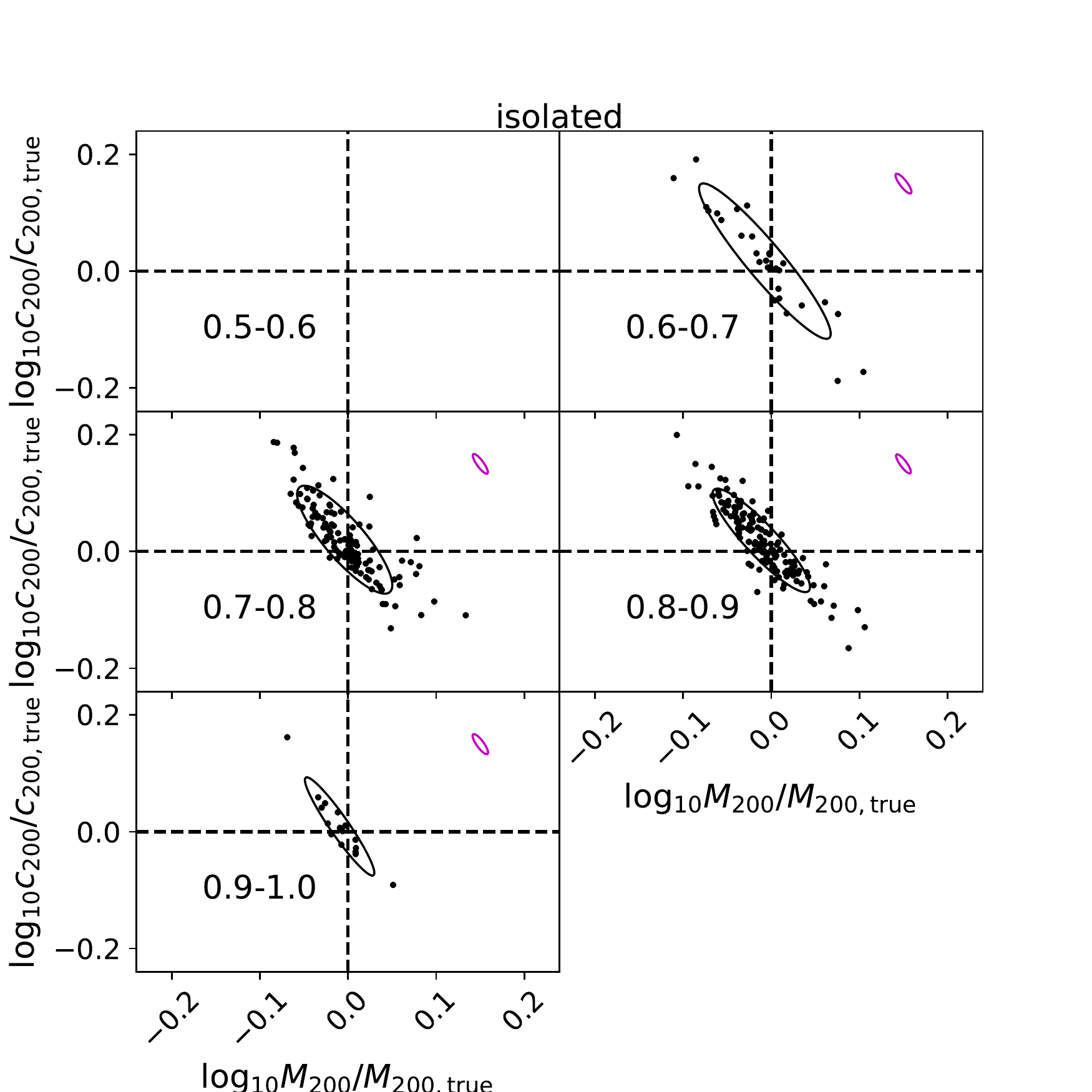}
\includegraphics[width=0.49\textwidth]{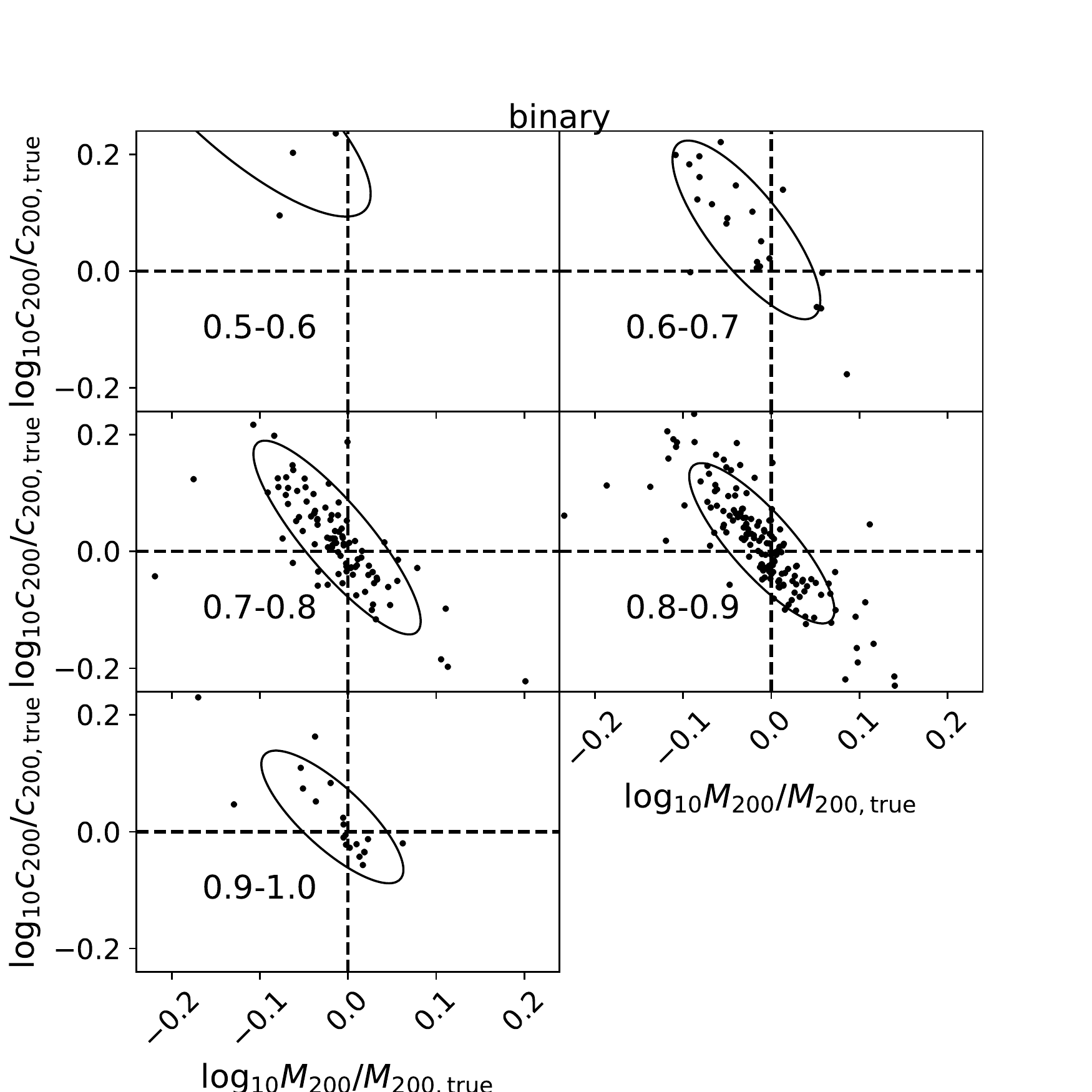}
\caption{Best-fitting concentration and halo mass for isolated (left) and binary 
(right) haloes in MRII, shown in separate panels according to the minor-to-major 
axis ratios of their inertia tensor within $R_{200}$, as labelled on each panel. 
The small magenta ellipse in each panel of the left plot shows the size of the 
statistical errors.} 
\label{fig:MRIIca} 
\end{figure*}

Results based on Equation~\ref{eqn:pot} are shown in Fig.~\ref{fig:MRIIcomb}. The 
left plot is for all selected isolated and binary haloes as a whole. Similar 
to \cite{2017MNRAS.470.2351W}, the 1-$\sigma$ uncertainty in best fits (black ellipse) 
is calculated and plotted by excluding the most biased measurements using 
3-$\sigma$ clipping, that is, to first calculate the 1-$\sigma$ uncertainty 
using all measurements that have converged and then estimate the uncertainty 
again by excluding data points outside 3 times the size of the 1-$\sigma$ 
uncertainty. The result indicates a scatter of about 25\% in $M_{200}$ and 
40\% in $c_{200}$. 

The uncertainty is close to that of oPDF in \cite{2017MNRAS.470.2351W} 
(see the red ellipse in the middle). Although the detailed approaches of oPDF 
and the SJE are different, they are both based on the steady state  
and spherical assumptions, and thus we should expect comparable level of 
uncertainties. According to the Cramer-Rao theorem, maximum likelihood 
estimators constructed from the full distribution function should be the 
most efficient. As a result, we would naively expect the oPDF maximum 
likelihood estimator to give a smaller uncertainty than the SJE, which 
is based on the momentum of the distribution function. The oPDF predicts the 
radial distribution of tracers given a model potential and the current 
radius, radial and tangential velocities of tracers. However, as currently implemented, 
it obtains the best-fitting halo potential by comparing only the 
predicted and ``observed'' radial distributions of the
tracers, and thus not all the available information is used, which 
might explain why the oPDF gives slightly larger uncertainties.

In contrast with results based on the ideal tracer sample above, the statistical 
error is {\it much smaller} than the uncertainty in best-fitting halo parameters, 
indicating the existence of various systematics (see equation~\ref{eqn:errors}). 
The oPDF gives the same result. Despite the different sizes, both the statistical 
error and the uncertainty in best fits tend to align with an anti-correlation 
direction of $M_{200}$ and $c_{200}$. \cite{2017MNRAS.470.2351W} attributed this 
alignment to the existence of the first systematic error, ${\bf C}_{\rm steady}$, 
which is introduced by phase-correlated structures such as streams. 

The orbital phase of a particle is a measure of its location on a given orbit (or 
more precisely, its travel time on the orbit, see \citealp{han2016a,han2016b,2017MNRAS.470.2351W} 
for a more precise definition), while phase-correlation refers to the clustering 
of particles along the orbit, forming coherently moving structures such as streams. 
The time evolution of streams thus makes the phase space distribution function of 
the system evolve over time, violating the steady state assumption. Thus the error 
introduced by these structures are classified as ${\bf C}_{\rm steady}$. On the other 
hand, this violation can be largely accounted for if one properly considers the 
number of phase-independent particles, which is expected to be smaller than the 
actual number of tracer particles. In other words, ${\bf C}_{\rm steady}$ can be 
modelled as a statistical error determined by an effective number of phase independent 
particles. This explains why the error ellipse of the scatter takes similar shape 
as the statistical error, while the amplitude is larger. Despite the similar origin, 
\citet{2017MNRAS.470.2351W} has shown that this effective number is almost independent 
of the tracer sample size. Accordingly, ${\bf C}_{\rm steady}$ is a systematic 
uncertainty determined by the intrinsic property of the system. In Section~\ref{sec:ca} 
and Section~\ref{sec:steady} we will make further discussions on this interpretation 
after disentangling different sources of systematics.

The right plot shows in black and green symbols results based on binary and 
isolated haloes separately. Note the mass distribution of binary haloes are 
biased to be smaller than that of isolated haloes according to our selection 
criterion. In order to ensure the same halo mass distribution we have matched 
each binary halo to one isolated halo according to their $M_{200}$. This is 
to avoid the bias caused by possible correlations between halo dynamical 
properties and $M_{200}$. 

The uncertainty in best-fitting parameters for the binary population is 
slightly larger than that of the isolated population, in good agreement 
with \cite{2017MNRAS.470.2351W}. The reasons are twofold. First, binary 
haloes are more aspherical than isolated ones. Moreover, the dynamical status 
of binary haloes are more disturbed than isolated haloes due to the existence 
of a nearby massive companion. In the following section, we demonstrate these 
two effects separately.

\subsubsection{Deviations from spherical symmetry vs violations of steady state}
\label{sec:ca}

In order to separate the effect of deviations from spherical symmetry and violations 
of the steady state assumption, we divide haloes into different subsamples based on 
their minor-to-major axis ratio, $c/a$, which is computed from the inertia
tensor obtained from the mass distributions within $R_{200}$. Results are shown in 
Fig.~\ref{fig:MRIIca} for isolated and binary haloes separately. 

There is a clear trend for the uncertainty to increase with decreasing $c/a$. 
Interestingly, we can see clearly that for the binary population, the measurements 
are biased more and more towards the upper left region for the most elongated haloes. 
Note this is not seen in \cite{2017MNRAS.470.2351W}. Though oPDF also involves the 
spherical assumption, \cite{2017MNRAS.470.2351W} used the true underlying potential 
profiles as templates when fitting to isolate possible violations of the NFW model, 
which helped to reduce this effect. In this paper, we focus on the NFW model, but 
with the oPDF method, we have repeated the analysis using the NFW model and see a 
similar bias for those most elongated haloes.

There are no isolated haloes with $c/a<0.6$, indicating the existence of more 
elongated binary haloes. Moreover, for fixed $c/a$, the scatter for  binary haloes is
larger than for isolated ones. This suggests the larger scatter in the binary population 
is not only because binaries are more elongated, but also because the dynamical 
status of binaries are more disturbed, due to the gravitational influence of their 
companion haloes. 

MW and M31 form a binary system, and it is interesting to see where our MW 
lies in Fig.~\ref{fig:MRIIca}. \cite{2013ApJ...773L...4V} measured its minor to 
major axis to be 0.8, which is slightly oblate. It sits in the panel where the 
uncertainty in best-fitting halo parameters is larger than the most spherical 
haloes, but the average measurement is close to being ensemble unbiased.

The statistical error is almost independent of the halo shape for isolated 
haloes. 
For the most spherical haloes with $c/a>0.9$, since violations of the spherical assumption 
(${\bf C}_{\rm sph}=0$) and deviations from the NFW profile (${\bf C}_\mathrm{NFW}\approx0$) 
are negligible, the remaining source of systematic error is violations of the steady state 
assumption, i.e., ${\bf C}_{\rm best-fit}\approx {\bf C}_{\rm steady}+{\bf C}_{\rm stat}$. 

The statistical error of these most spherical haloes is still about 5 times smaller than 
the uncertainty in best fits. If one properly considers the effective number of phase-independent 
particles, ${\bf C}_{\rm best-fit}$ is expected to be purely statistical. Since the statistical 
error scales with sample size $N$ as $\sigma^2 \propto 1/N$, we can have an estimate of the 
number of phase-independent particles as 
\begin{align}
N_{\rm eff}&=N_{\rm tracer}\frac{\sigma^2_{\rm stat}}{\sigma^2_{\rm best-fit}}\\
&=10^5\times\left(\frac{1}{5}\right)^2\nonumber\\
&=4000.\nonumber
\end{align}
This is in good agreement with the number estimated in \cite{2017MNRAS.470.2351W}.

\subsubsection{Reducing phase-correlated dynamical tracers}
\label{sec:steady}

\begin{figure} 
\includegraphics[width=0.49\textwidth]{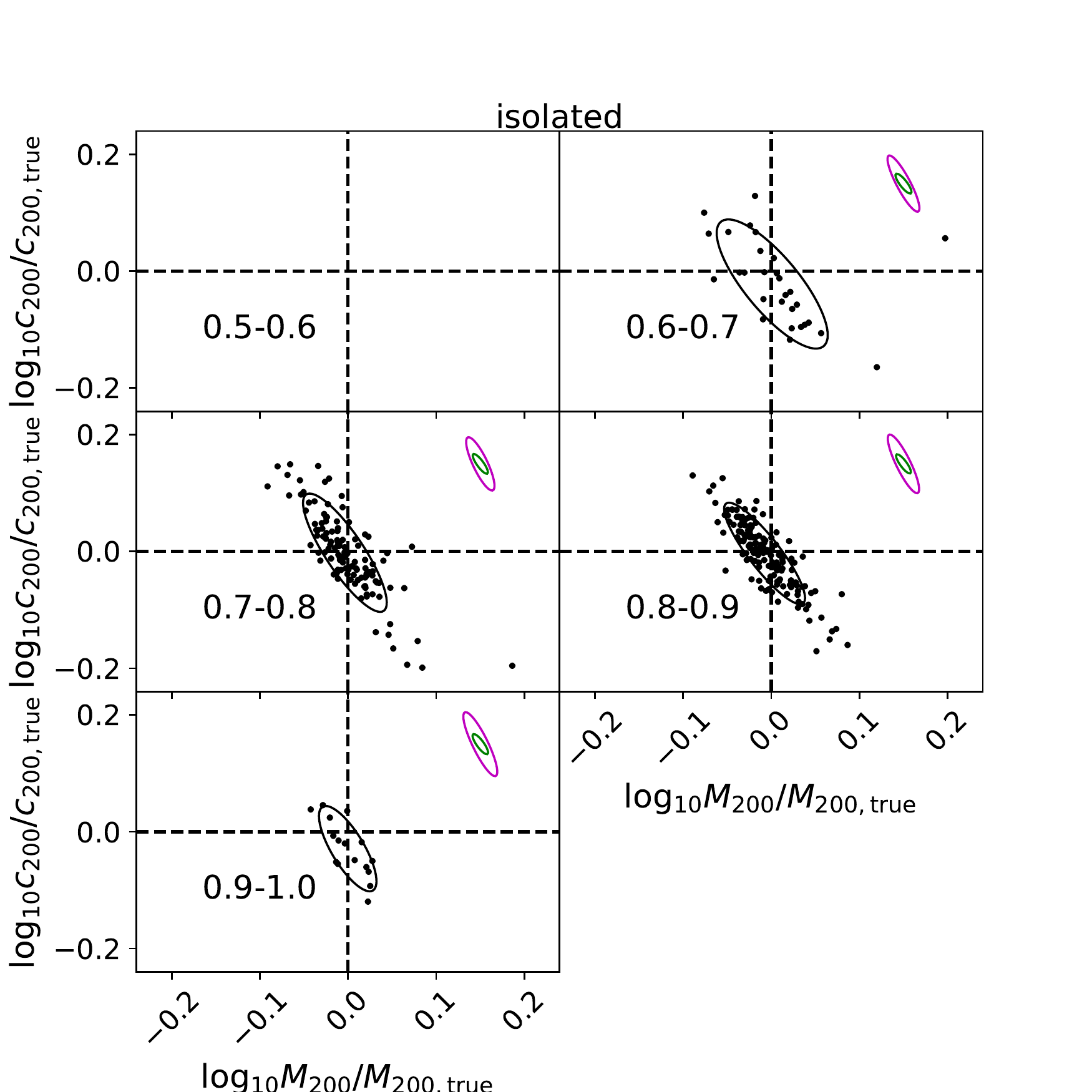}
\caption{Similar to the left plot of Fig.~4, but each particle has been weighted 
by the inverse of the total number of particles in the stream to which it belongs. 
Streams are identified by tracing particles back to their progenitors using 
the \hbtp code of Han et al. (2012, 2017). Magenta ellipses are the new 
statistical errors after weighting, whereas green ellipses repeat the statistical 
errors from the left plot of Fig.~4. } 
\label{fig:MRIIca_weight} 
\end{figure}

The argument of under-estimated statistical errors due to phase-correlated 
structures can be tested in the following way. The basic idea is to degrade 
the sample size by down weighting phase-correlated particles. We implement 
this in an extreme way, by treating particles from the same progenitor as 
being in a phase-correlated stream, and restricting each stream to contributing 
only one degree of freedom. After that we can check whether the statistical 
error is then closer to the uncertainty in best fits, at least for the most 
spherical haloes.

We start by tracing particles back in time to identify their progenitors, 
using the halo merger tree built by the \hbtp code
\citep{2012MNRAS.427.2437H,2017arXiv170803646H}. \hbtp normally does unbinding 
to remove unbound particles during halo tracking. For the purpose of this study, 
we have switched off unbinding in \hbtp to allow a complete recovery of 
phase-correlated particles. Particles once belonging to a progenitor at 
its maximum mass are treated as coming from this progenitor. Note that hierarchical 
merging of progenitor haloes may lead to the formation of subhalo groups (or 
groups of streams in our case), which move coherently inside the host halo. Such 
cases can potentially cause correlations even among different streams, which is 
hard to account for in our current analysis.

We label the total number of particles in stream i as $N_i$, and each particle 
is assigned a weight of $1/N_i$. Smoothly accreted diffuse particles that do not 
belong to any progenitor haloes are assigned weights of 1. We repeat our analysis 
in section~\ref{sec:method} with this weighted sample. The weights can be thought 
as the ``mass'' associated with each particle, which helps to decrease the 
contribution from massive streams, whose particle number is large but the effective 
number of phase-uncorrelated particles can be much smaller.

Note, however, this weighting scheme ignores the internal structure of streams. 
In reality, different streams may contribute different number of phase-independent 
particles, since the correlation strength among stream particles depends on factors 
such as the infall time and orbit. Particles in streams accreted earlier have longer 
time to reach a more phase-mixed status, and hence may contribute a larger number of 
phase-independent particles. Moreover, with this weighting, the SJE still holds only 
if all streams are independent populations. This is not necessarily true due to 
possible correlations among streams mentioned above. So our test is a simplified approach.

Results are presented in Fig.~\ref{fig:MRIIca_weight}. We only show isolated 
haloes for a clean picture, as for binaries the dynamical status is affected by the 
massive companion. Compared with the left plot of Fig.~\ref{fig:MRIIca}, the uncertainty 
in best-fitting halo parameters (${\bf C}_\mathrm{best-fit}$) is almost the same, and perhaps 
only slightly decreased along the major axis of the error ellipse. This suggests that 
down-weighting stream particles has not led to any significant loss of dynamical information. 
The statistical errors are inferred from bootstrap sampling as before except that each 
particle is assigned a weight according to its original stream mass. There is a small 
change in the shape and direction of the statistical error ellipse, due to the introduction 
of weights. The error ellipse of best fit parameters also changes correspondingly.  
Encouragingly, the new statistical errors after weighting are significantly larger and 
become comparable to the uncertainty in best fits for the most spherical bin. Since 
${\bf C}_\mathrm{best-fit}={\bf C}_\mathrm{steady}+{\bf C}_\mathrm{stat}$,  
this means the systematic caused by phase-correlated particles that violates the steady 
state assumption, ${\bf C}_{\rm steady}$, is much reduced. This can be understood as 
contributions from phase-correlated particles are suppressed in bootstrap due to their 
lower weights, enabling the bootstrap process to correctly sample the variations from 
phase-independent particles, thus capturing the true degree of freedom of the system.

The test leads support to our argument that phase correlations in streams 
violate the steady state assumption and cause underestimates of the statistical errors. 
But we should note various factors that can affect the statistical error size. If the 
internal structure of streams can be estimated, the statistical error would be smaller. 
Moreover, correlations between streams mentioned above, if considered, would increase 
the statistical error. Lastly, smoothly accreted particles might also be phase-correlated 
due to coherent infall along large cosmic filaments, corresponding to `unresolved' streams. 
Taking streams which can not be resolved into account, the statistical error size is 
expected to further increase.

\subsection{Stellar tracers in APOSTLE}
\label{sec:LG}

\begin{figure} 
\includegraphics[width=0.49\textwidth]{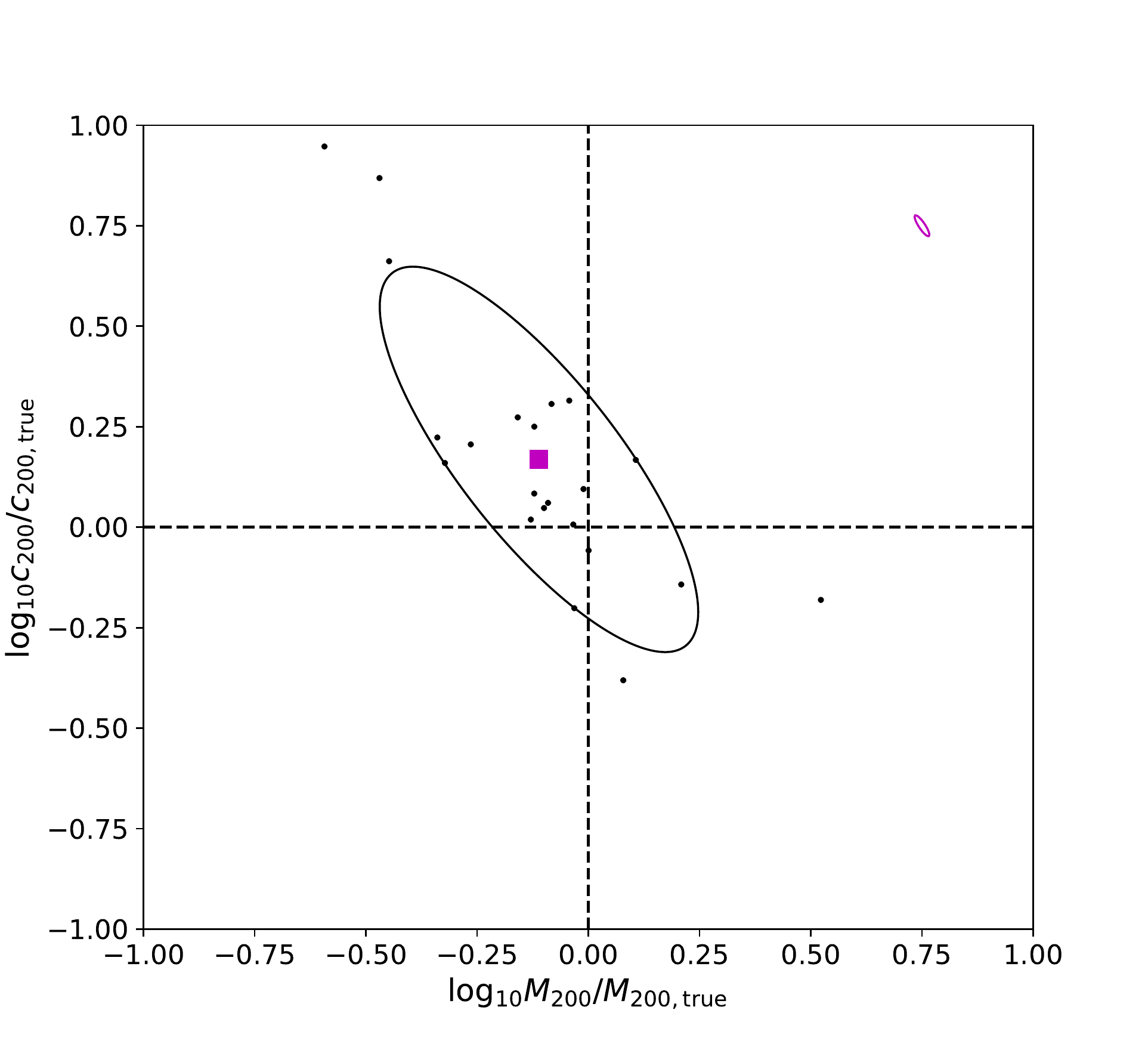}
\caption{Best-fitting halo parameters for galaxies in the APOSTLE simulations. 
Points are fits to individual galaxies and the black ellipse marks the 1-$\sigma$ 
scatter. The magenta solid square is the average parameter of all the galaxies. 
The tiny magenta ellipse in the top right corner shows the 1-$\sigma$ statistical 
errors averaged over the entire sample of galaxies. An inner radius cut of 
20~kpc has been adopted to exclude the disc component.} 
\label{fig:2Dapostle}
\end{figure}

The large sample of MRII haloes enables us to compare the systematic uncertainty 
with statistical noise and investigate the dependence on the halo shape. However, 
dark matter particle tracers are not representative of real observations. 
In the following we further test the model performance by using stellar tracers 
in the APOSTLE simulations, in order to have results more closely related  
to the real observations. 

Results are presented in Fig.~\ref{fig:2Dapostle}. Note the axis range is 
different from that in previous figures. One can see a much larger uncertainty 
in the best-fitting parameters, which is about a factor of three in both mass 
and concentration. 
The statistical error is again much smaller than the uncertainty in best-fitting 
halo parameters. The true effective number of phase independent particles is 
estimated to be around only 40. However, due to the small sample size, it is 
difficult for us to separate contributions from violations of the spherical 
assumption, and thus this number of 40 should only be regarded as a lower limit. 

The larger uncertainty and smaller number of phase independent particles for 
stars can be understood in the following way. First of all, stars in 
the stellar halo are usually believed to be stripped from a small number of 
satellite galaxies, which are hosted by dark matter subhaloes and are the most 
bound part of those subhaloes. Compared with dark matter particles in these subhaloes, 
stars are stripped late due to their higher binding energy and thus they have 
less time to reach a steady state. This introduces stronger violations of 
the steady state assumption, and it is natural to expect a smaller number of 
phase independent particles. Moreover, dark matter particles in the host halo are 
not only formed through stripped particles from substructures, but also through 
smooth accretion of background particles. These smoothly-accreted particles 
are expected to be more relaxed and phase independent \citep{2011MNRAS.413.1373W}.
Lastly but importantly, baryonic physics such as mass ejections produced by 
supernova explosions and stellar winds, can act to violate the steady state assumption of 
stellar tracers. This process does not exist in dark matter only simulations.

The overall uncertainty in Fig.~\ref{fig:2Dapostle} is in good agreement with 
\cite{2017MNRAS.470.2351W}, but the measurements are, on average, slightly biased 
towards the upper left corner. Such an apparent overall bias might be due to the 
small sample of 24 APOSTLE galaxies. A larger sample might give results that are 
less ensemble biased. On the other hand, it might indicate some systematics. 
\cite{2017MNRAS.470.2351W} 
have discussed that if using the NFW potential profile to model the underlying 
potential in hydro-dynamical simulations, the result would be ensemble biased 
in such a direction. The bias is more obvious with more particles in the inner 
region included. In fact, \cite{2015MNRAS.446..521S} found that the presence of 
stars can produce cuspier inner profiles than the NFW model, and the effect is most 
prominent in haloes of masses about $10^{12}$ to $10^{13} \msun$. Since we have 
excluded particles within 20~kpc, the bias is unlikely to be due to the deviation 
from the NFW model in inner regions. It is more likely due to the deviation from 
the spherical assumption (see Fig.~\ref{fig:MRIIca}), as all APOSTLE galaxies are 
in pairs and the underlying potential profiles are more likely to be more elongated.

\section{Discussion}\label{sec:discussion}

\subsection{Constant or improper $\beta$?}
\label{sec:fixbeta}

\begin{figure*} 
\includegraphics[width=0.49\textwidth]{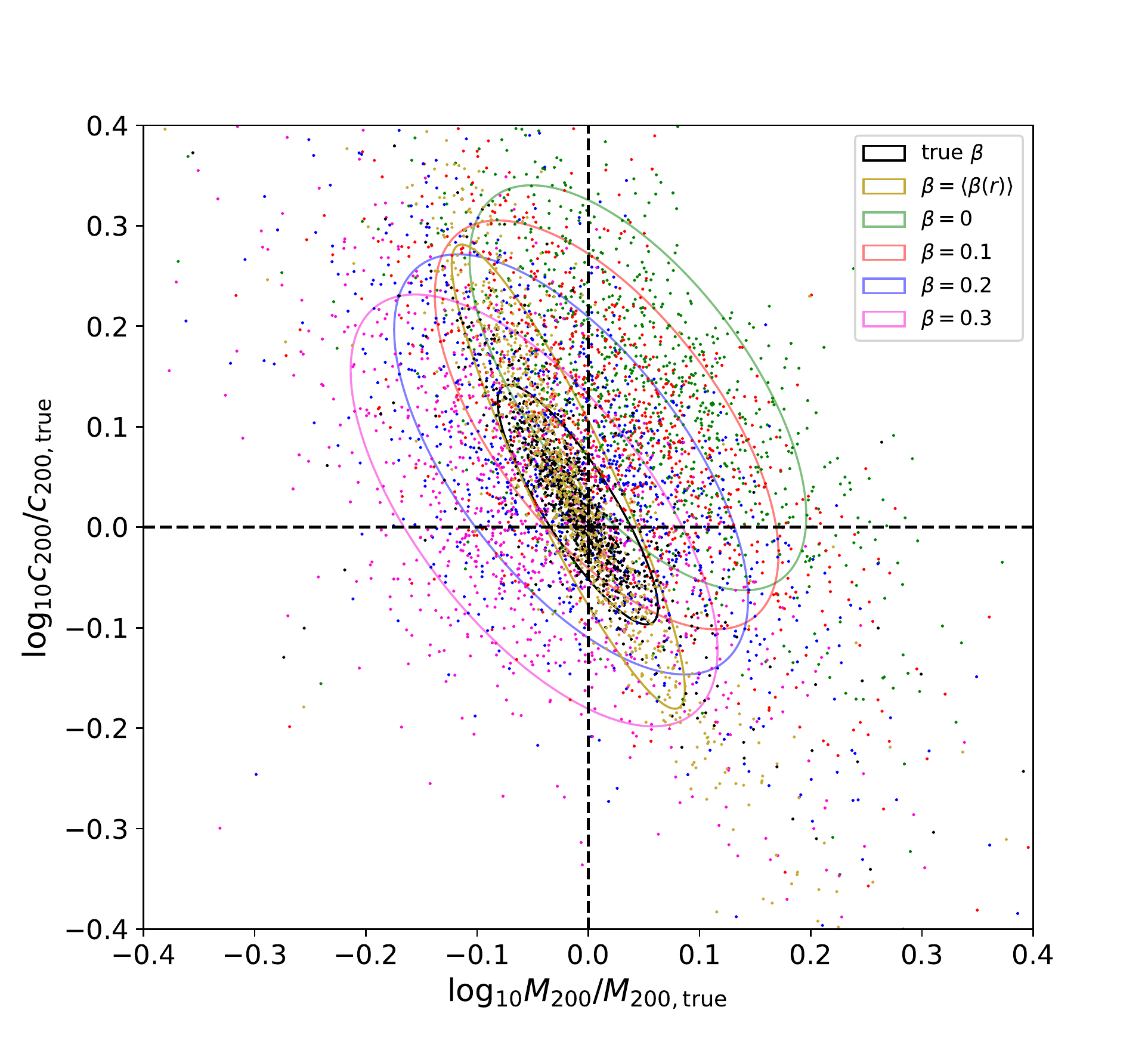}
\includegraphics[width=0.49\textwidth]{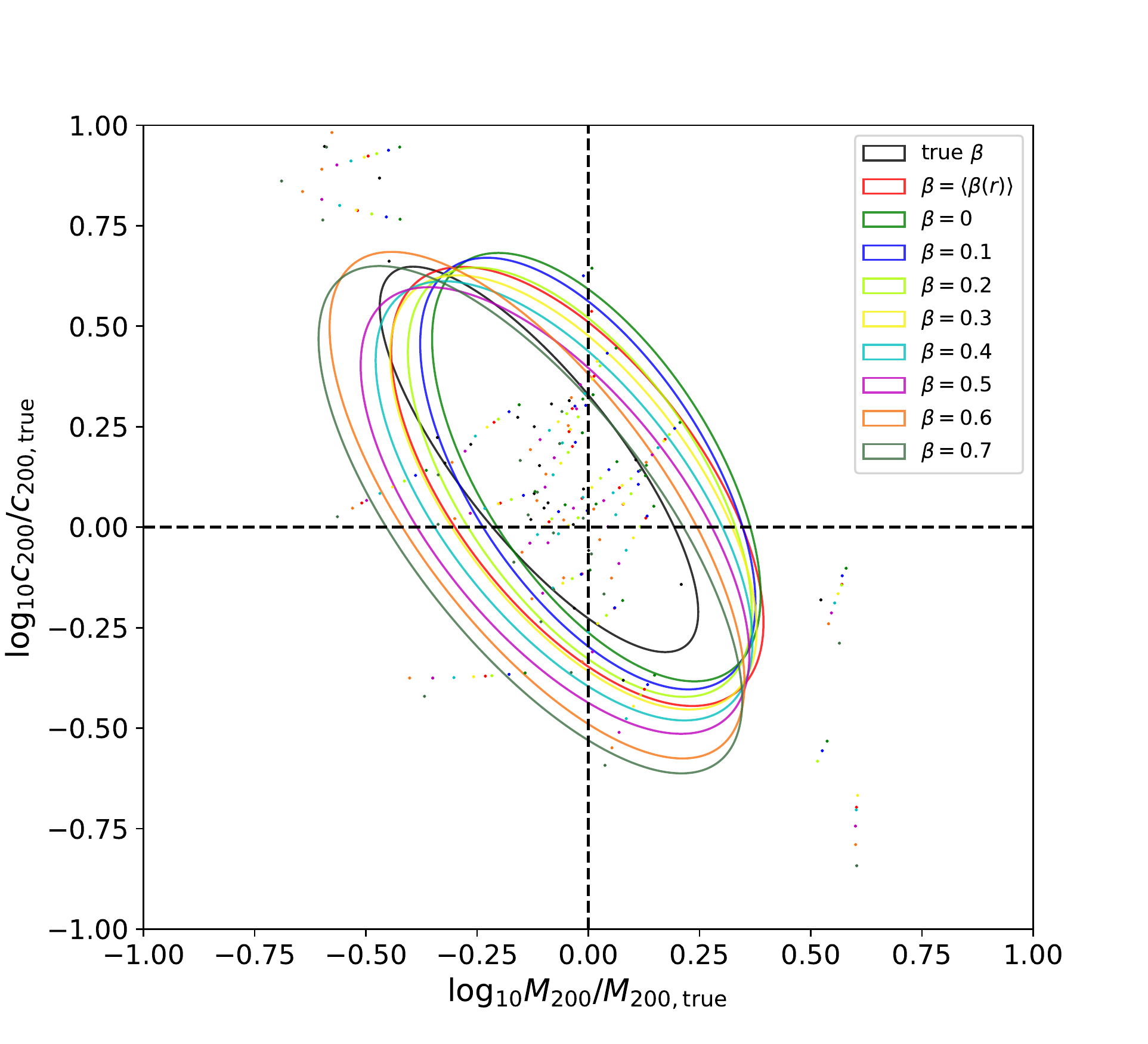}
\caption{Best-fitting halo parameters when, instead of using the true $\beta(r)$ profiles 
in the simulation, $\beta$ is assumed to take a fixed values over the whole radial range. 
Dots and the 1-$\sigma$ uncertainty ellipses are colour coded by the value of $\beta$, 
indicated in the legend. $\langle \beta(r) \rangle$ is the mean value of $\beta(r)$ averaged 
over $20~{\rm kpc}<r<R_{200}$ and varies for individual haloes. Calculations are shown 
for MRII (left) and APOSTLE (right). Black dots and ellipses in each panel repeat the 
results from Fig.~\ref{fig:MRIIcomb} and Fig.~\ref{fig:2Dapostle} respectively to enable 
direct comparisons. } 
\label{fig:beta}
\end{figure*}

In reality, the tangential velocities of tracers are often missing. As a result, the 
velocity anisotropy of tracers, $\beta$, cannot be measured directly. To get over 
this issue, one has to either adopt some fiducial values of $\beta$, or assume a 
certain parametrised form of $\beta(r)$, and get the best-fitting parameters. 
In this subsection, we first test the effect of fixing $\beta$ in the left side of 
Equation~\ref{eqn:pot} to a given value and using the NFW model potential to fit 
the inferred potential gradient. In the next subsection, we further treat $\beta$ 
as a free parameter.

Results are presented in Fig.~\ref{fig:beta}, where we fix $\beta$ to a few pre-set values.
From $\beta=0$ to larger positive $\beta$, the results gradually change from over-estimated 
$M_{200}$ ($c_{200}$) to under-estimated $M_{200}$ ($c_{200}$) on average.  This can be 
understood as the degeneracy between $\beta$ and $\ud \Phi/\ud r$ in Equation~\ref{eqn:pot}: 
to maintain the first term on the left side unchanged, increased $\beta$ corresponds to 
decreased $\ud \Phi/\ud r$. The scatter remains almost unchanged with different $\beta$, 
and is slightly larger for very large (tangential) values of $\beta$ in APOSTLE. This 
scatter is larger than previous results that used the true $\beta(r)$ profile. 

We also test the case when each halo adopts the average $\beta$ of its tracer sample, 
$\langle \beta(r) \rangle$. The scatter is reduced perpendicular to the mass-concentration 
anti-correlation direction, because each halo adopts its own average $\beta$ rather than 
adopting a common pre-set value. However, the scatter along the anti-correlation direction 
remains larger than the previous result adopting the true $\beta(r)$ profile. Encouragingly, 
the result becomes mostly ensemble unbiased.

\subsection{Radius-independent $\beta$ as a free parameter?}

\begin{figure*} 
\includegraphics[width=0.49\textwidth]{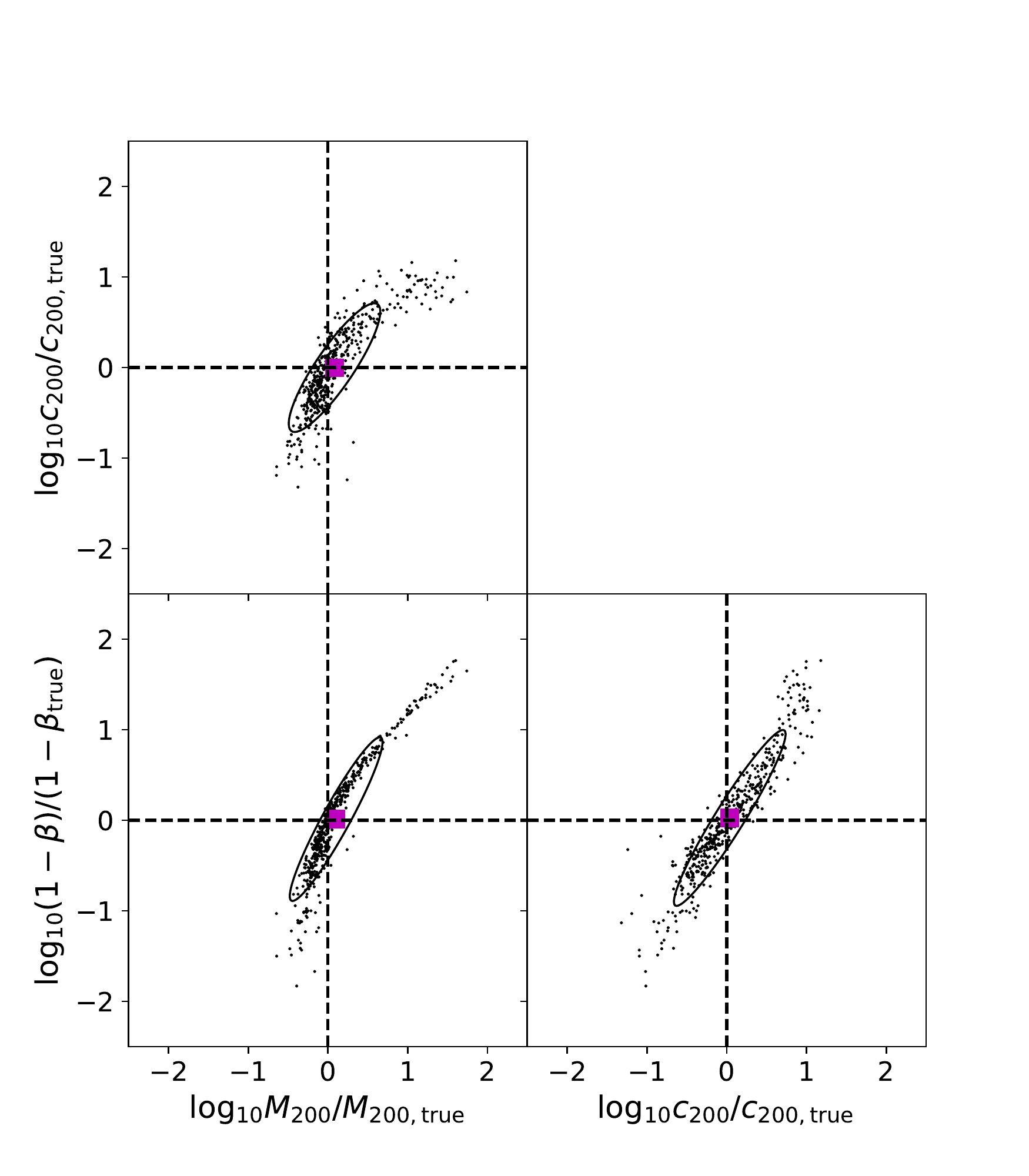}
\includegraphics[width=0.49\textwidth]{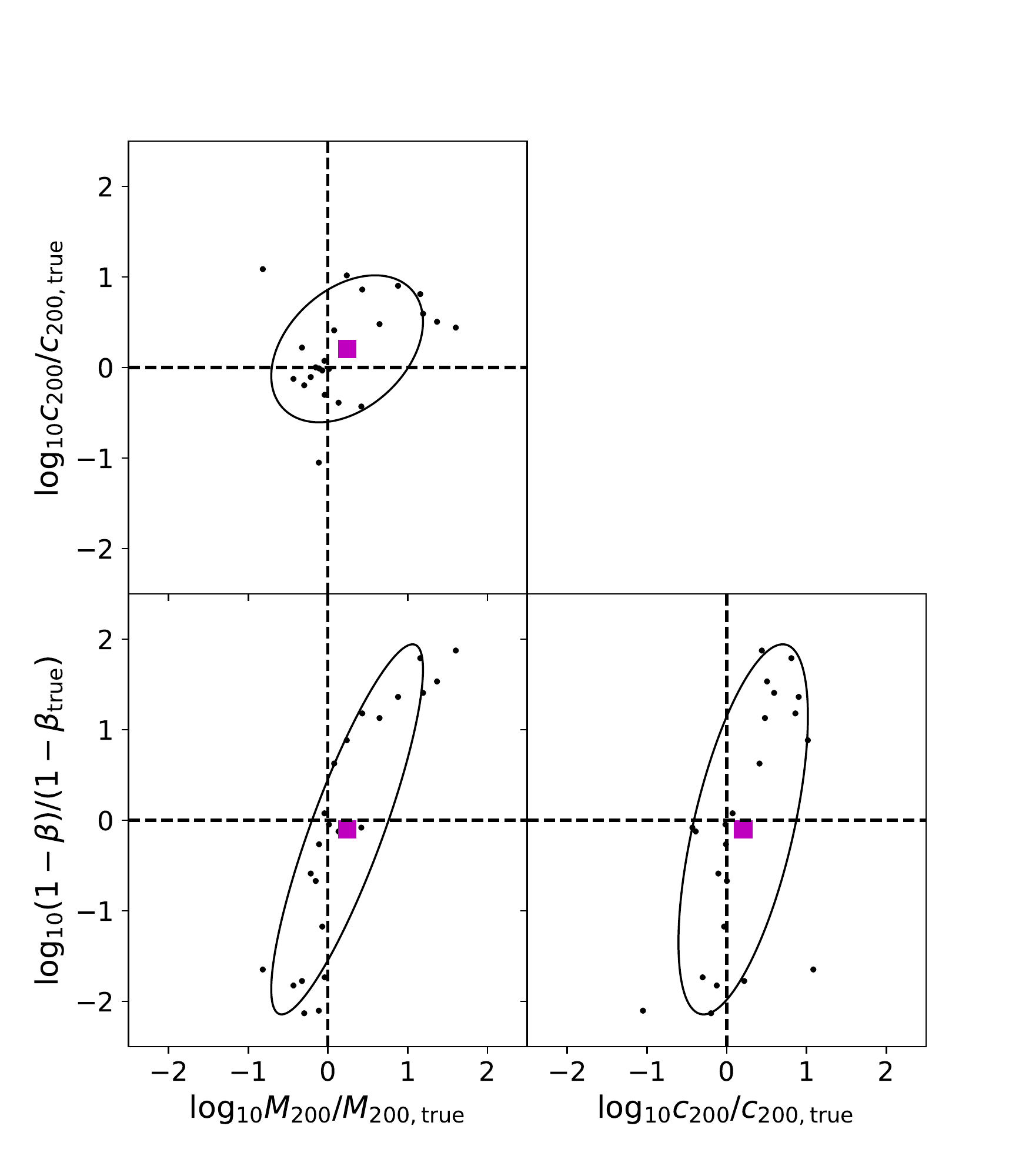}
\caption{Best-fitting $M_{200}$, $c_{200}$ and $\beta$ for isolated haloes in MRII 
(left) and galaxies in APOSTLE (right) using Equation~\ref{eqn:sigmar}. The fits are 
made to the measured radial velocity dispersion profiles of tracers, while $\beta$ is 
treated as a radius-independent free parameter. The black ellipse and magenta 
square in each panel mark the 1-$\sigma$ uncertainty and the average measurement. } 
\label{fig:betafree}
\end{figure*}

We now directly fit the measured radial velocity dispersion profiles from the 
simulations using Equation~\ref{eqn:sigmar} by treating $\beta$ as a constant but 
free parameter. This is a common approach adopted in the literature 
\citep[e.g.][]{2005MNRAS.364..433B,2014ApJ...794...59K}. Note it is 
different from the approach in Section~\ref{sec:fixbeta}, where $\beta$ was 
fixed to an predefined constant value and the potential gradient inferred from 
Equation~\ref{eqn:pot} was used for fitting. 

Fig.~\ref{fig:betafree} shows results for both isolated MRII haloes and APOSTLE galaxies. 
Note we focus on isolated haloes in MRII, because the radial density profiles 
of dark matter tracers are affected by the massive companion halo in binary systems. 
This makes the integral of Equation~\ref{eqn:sigmar} hard to converge. The problem 
does not exist when stars in APOSTLE are used as tracers, because the radial 
density profiles of stellar tracers drop quickly on large distance. 

The axis range is larger than all previous figures. The results end up with much 
larger uncertainties, which can be as large as one order of magnitude, whereas 
for APOSTLE galaxies the uncertainty of $\beta$ can be two orders of magnitude. 
The larger scatter in $\beta$ for APOSTLE is probably because the velocity anisotropy 
of stars depends more strongly on radius than that of the dark matter, and thus 
modelling $\beta$ as a free but radius-independent parameter introduces larger systematic 
errors.

Compared with the large scatter, the measurements averaged over all haloes/galaxies 
are much closer to zero. There are about 0.1 dex of deviation in the average $M_{200}$ 
towards the positive direction and in the average $\beta$ towards the negative direction.

$\beta$ shows strong anti-correlations with both $M_{200}$ and $c_{200}$. This 
trend has already been revealed in Fig.~\ref{fig:beta} that increased $\beta$ 
corresponds to underestimated $M_{200}$ and $c_{200}$. Interestingly, marginalising over 
$\beta$ leaves positive correlation between $M_{200}$ and $c_{200}$, in contrast 
with the anti-correlation found previously. This means the error is mainly driven 
by the uncertainty in $\beta$: a large deviation in $\beta$ from the true value 
leads to large deviations in both $M_{200}$ and $c_{200}$ along the same direction, 
resulting in the positive correlation between $M_{200}$ and $c_{200}$. This also 
leads to the comparable size of scatter in the best-fit $M_{200}$ and $c_{200}$ 
parameters between MRII and APOSTLE. This is in contrast to previous results 
that APOSTLE galaxies show significantly larger scatters than MRII haloes, when 
true $\beta$ profiles and Equation~\ref{eqn:pot} are used.

In Fig.~\ref{fig:MRIIcomb} and Fig.~\ref{fig:MRIIca}, not only the average 
statistical error aligns with the uncertainty in best-fits, but also above 98\% 
haloes show an anti-correlation between $M_{200}$ and $c_{200}$ in their 
individual statistical error. In contrast to Fig.~\ref{fig:MRIIcomb}, now the 
statistical errors show significant halo-to-halo scatter and no longer align with 
the uncertainty in best-fitting halo parameters. Due to the large scatter, it is 
not straight-forward to compute the average statistical error, which is hence not 
directly shown in Fig.~\ref{fig:MRIIcomb}.

We compare measurements within and outside the 1-$\sigma$ uncertainty ellipse in 
the corresponding parameter plane, and find that for measurements with larger 
deviations, the radial dependence of $\beta$ is stronger. On average, the $\beta$ 
profile for measurements outside the 1-$\sigma$ uncertainty ellipse drops faster 
at $r>100$~kpc by about 30\%. However, the halo to halo scatter of the $\beta$ 
profile is very large, and thus we avoid over-interpreting the result.

\subsection{Radial range of tracers}

\begin{figure*} 
\includegraphics[width=0.49\textwidth]{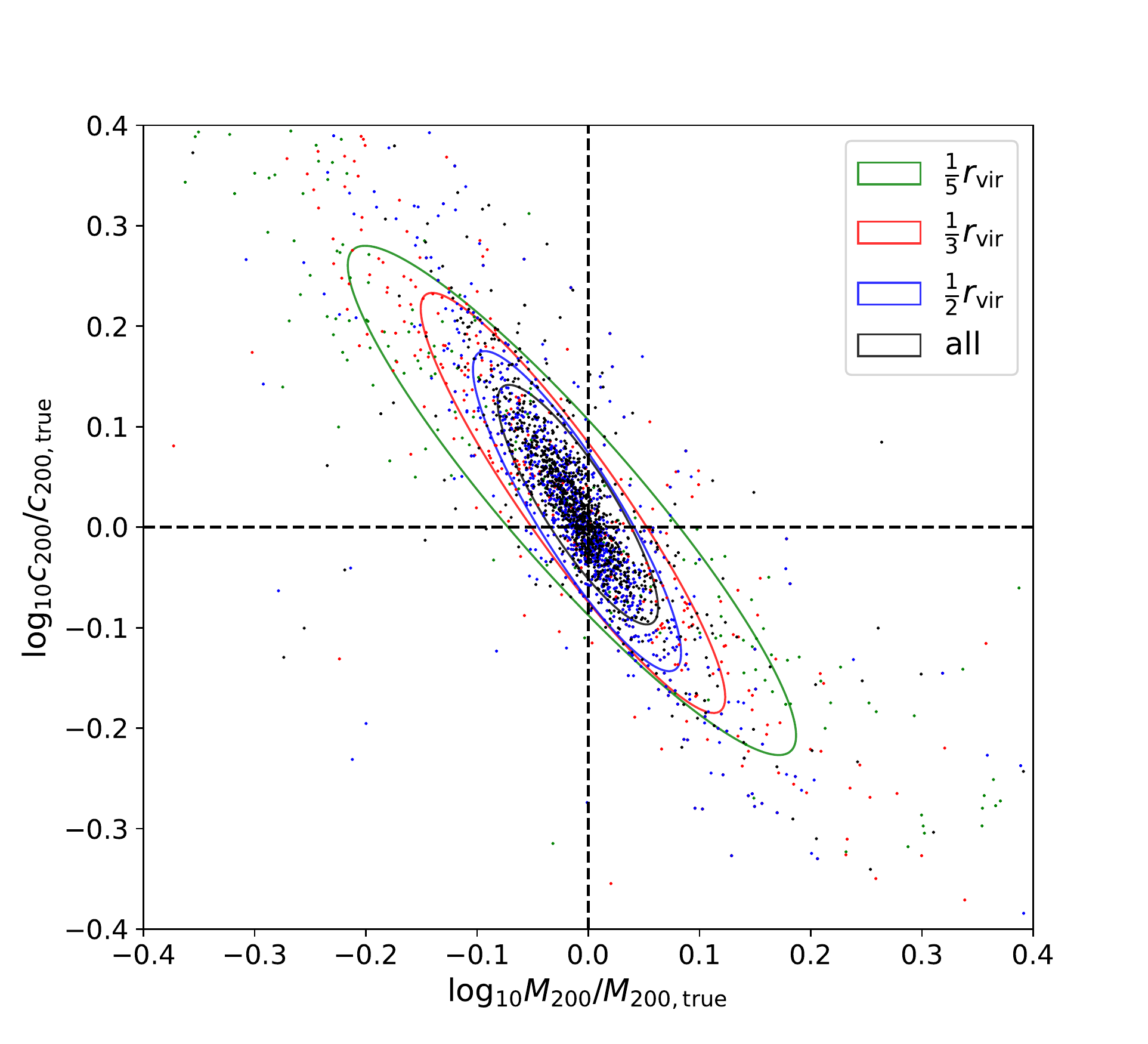}
\includegraphics[width=0.49\textwidth]{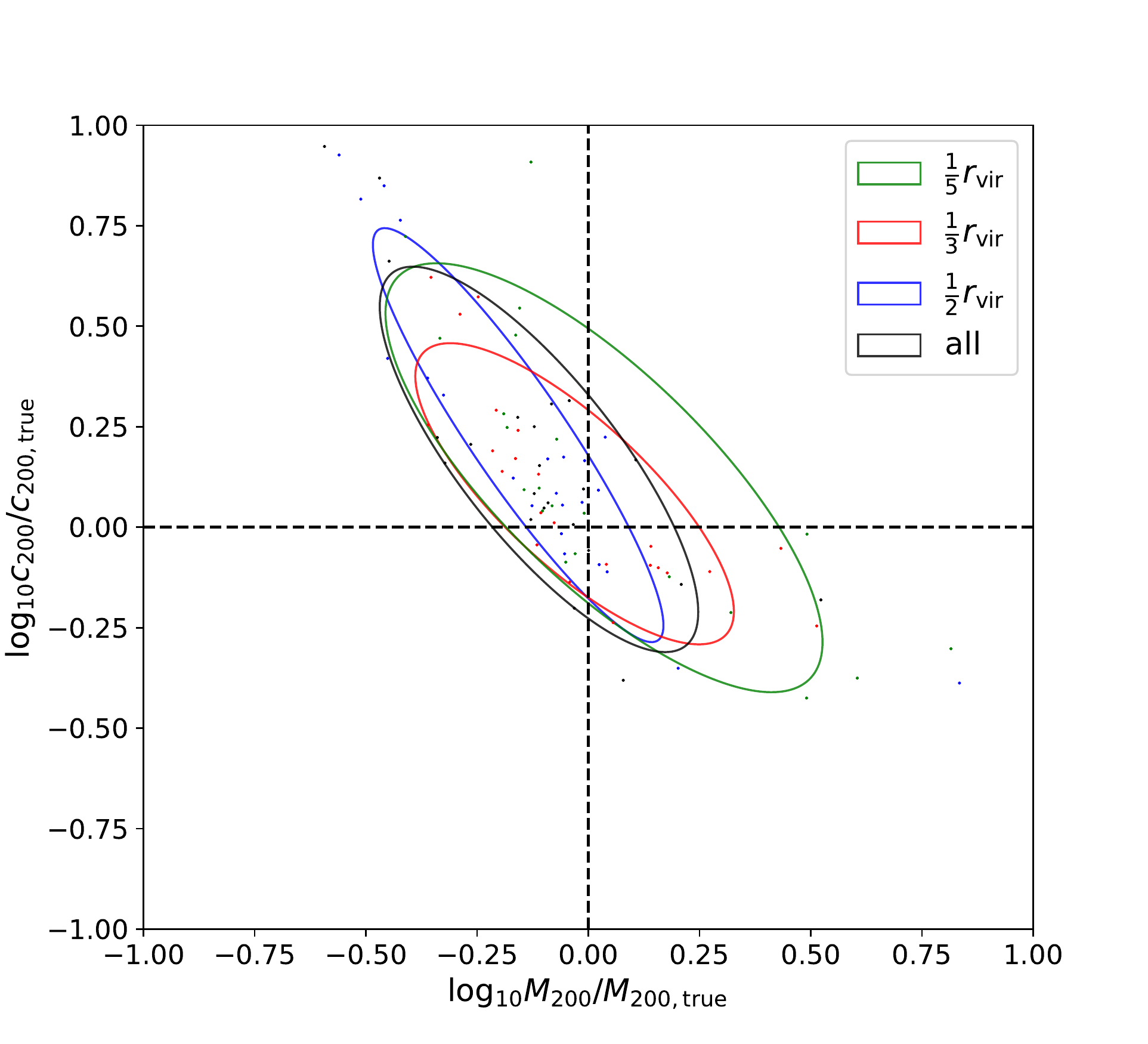}
\caption{Best-fitting $M_{200}$ and $c_{200}$ when using tracers within a fixed 
outer radius cut. Each dot represents one halo/galaxy. Measurements are colour 
coded by the choice of this outer radius, indicated in the legend. Calculations 
are shown for MRII (left) and APOSTLE (right). } 
\label{fig:rcut}
\end{figure*}

We are interested in the question of what will happen if only tracers within 
a limited radial range are used. This has been briefly discussed in 
\cite{2015MNRAS.453..377W}, where it was found that the total mass within 
the half-mass radius of all tracers over the whole radial range can be well 
constrained even when only a subsample of tracers within 60~kpc are used. However, 
with just five haloes, it is not very clear to see how $M_{200}$ is affected. 
Here we investigate this using both MRII and APOSTLE haloes and adopt 
three different radial cuts of 
$r<\frac{1}{5}R_{200}$, $r<\frac{1}{3}R_{200}$ and $r<\frac{1}{2}R_{200}$. 
We stick to Equation~\ref{eqn:pot} instead of Equation~\ref{eqn:sigmar}, in 
order to separate uncertainties due to improper modelling of $\beta$. 

Results are presented in Fig.~\ref{fig:rcut}. It is clear that with the reduction 
in tracer outer radius, the measurements show larger and larger overall scatter. 
The trend is not monotonic for APOSTLE, which might be due to the small 
sample size. Interestingly, although the scatter is significantly increased for 
$r<R_{200}/5$, the measurements are still close to be ensemble unbiased,
indicating the SJE and the NFW model profiles give good extrapolations to larger 
radii. 

\section{Conclusions}
The spherical Jeans equation (SJE) has been widely used to probe the mass 
profile or circular velocity curve of our MW galaxy. In this study we apply 
the SJE to more than 1000 dark matter haloes in the Millennium-II simulation 
and 24 MW-like galaxies in the APOSTLE simulations to investigate 
the performance of the SJE in recovering the halo potential, which we model 
as an NFW profile. 

The large sample of haloes and galaxies enables us to test the model in a 
statistically robust way. The best-fitting halo mass, $M_{200}$, and 
concentration, $c_{200}$, suffer from about 25\% and 40\%  systematic 
uncertainties when dark matter particles in MRII are used as dynamical 
tracers. When star particles in APOSTLE are used as tracers, the 
uncertainty can be as large as a factor of three. These uncertainties 
warn us that inferences based on a single case, such as our MW, are 
dangerous to make without quoting the large systematic errors behind. 
The uncertainties are in good agreement with the results of \cite{2017MNRAS.470.2351W}.
Although the detailed modelling approach of the oPDF used by 
\cite{2017MNRAS.470.2351W} is different from SJE, both the oPDF and SJE 
assume only steady state tracers and spherical potentials, and thus it is 
encouraging to see such a good agreement. The systematic uncertainty is 
attributed to violations of both the steady state and spherical symmetry 
assumptions. Any dynamical model relying on the two assumptions will 
suffer from a similar level of uncertainty. 

Haloes with minor to major axis ratio less than 0.7 have larger uncertainties 
and are, on average, biased towards underestimates in $M_{200}$ and 
overestimates in $c_{200}$. The latter is due to deviations from the NFW model, 
which can be eliminated if the exact density profiles are used as templates 
for the fitting \citep{2017MNRAS.470.2351W}. Binary haloes on average 
are more elongated and more disturbed than isolated haloes and show larger 
uncertainties in the fitting. 

We further verify the conclusion of \cite{2017MNRAS.470.2351W} that 
the statistical errors are underestimated because of phase-correlated structures 
such as streams. Assuming the statistical errors and the uncertainty in the 
best-fitting halo parameters would have the same size if one properly considered 
the true number of phase independent particles for the most spherical haloes
\footnote{By choosing the most spherical haloes, the systematic scatter 
is dominated by violations of the steady state assumption due to phase-correlated 
structures, and thus we are able to isolate the effect due to violations of the 
spherical symmetry assumption.}, the effective number of phase independent tracer 
particles is estimated to be about 4000 for dark matter (20~kpc inner cut). 
For stars, the lower limit is about 40. Such systematic uncertainties cannot be 
trivially reduced by simply increasing the total sample size, since they are 
determined by the intrinsic effective number of phase independent particles 
rather than the total number of particles. We can call this the {\it limiting 
precision} of dynamical modelling of the MW stellar halo. 

There are ways to decrease the number of phase-correlated particles. If one 
can exclude tracers from the few most massive and prominent streams, the phase-correlations 
would be reduced. In our analysis, we weight particles from the same stream by the 
inverse of the total particle number in this stream. This helps to down-weight the 
contribution from massive streams and brings closer agreement in the size of 
statistical errors and uncertainties in best-fitting halo parameters. 

Finally, we investigate the effect of improper modelling with a 
radius-independent $\beta$ parameter and the effect of only using 
tracers within a given radial range. Treating $\beta$ as a radius-independent 
but free parameter, we end up with much larger uncertainties in best-fitting 
halo mass, concentration and $\beta$ itself. The uncertainty can 
be as large as one order of magnitude but the average measurements 
are approximately ensemble-unbiased compared with the large scatter. 
Under-estimating $\beta$ causes overestimates in $M_{200}$ and $c_{200}$ 
and vice versa. When $\beta$ is a free parameter, $M_{200}$ positively 
correlates with $c_{200}$, as the uncertainty becomes primarily {\it driven by}
the error in $\beta$. 
Using tracer particles within a given outer radius can significantly increase 
the uncertainty. Nevertheless, even when only tracers within $R_{200}/5$ 
are used, the best-fitting halo parameters are almost ensemble unbiased, 
indicating the SJE and the NFW model profiles give good extrapolations to the outer 
radius of the dynamical system.

\section*{Acknowledgements}
Kavli IPMU was established by World Premier International Research 
Center Initiative (WPI), MEXT, Japan. This work was supported by JSPS 
KAKENHI Grant Number JP17K14271, and supported by the Science and Technology 
Facilities Council Durham Consolidated Grant [ST/F001166/1,ST/L00075X/1,ST/P000451/1]. 
The APOSTLE project used the DiRAC Data Centric system at Durham University, operated 
by the Institute for Computational Cosmology on behalf of the STFC DiRAC 
HPC Facility (www.dirac.ac.uk), and also resources provided by 
WestGrid (www.westgrid.ca) and Compute Canada (www.computecanada.ca).
The DiRAC system was funded by BIS National E-infrastructure capital 
grant ST/K00042X/1, STFC capital grants ST/H008519/1 and ST/K00087X/1, 
STFC DiRAC Operations grant ST/K003267/1 and Durham University. 
DiRAC is part of the National E-Infrastructure. WW is grateful for 
useful suggestions made by the referee for the first Wang et al. (2017) 
paper.

\bibliography{master}

\end{document}